\documentclass{aa}
\usepackage{graphicx}
\usepackage[euler]{textgreek}
\usepackage{breqn}
\usepackage{amssymb}
\usepackage{comment}
\usepackage{subfigure}
\usepackage{color}

\begin{document}


\title{Particle Acceleration and gamma-ray emission due to
magnetic reconnection around the core region of radio galaxies}
\titlerunning{Particle Acceleration and $\gamma$-ray emission by Magnetic reconnection}
\author{B. Khiali
          \inst{1}
          \and
           E. M. de Gouveia Dal Pino
          \inst{1}
          \and
          H. Sol\inst{2}
          }

\institute{IAG-Universidade de S\~ao Paulo, Rua do Mat\~ao 1226, S\~ao Paulo, SP, Brazil\\
              \email{bkhiali@usp.br}
         \and
             LUTH, Observatoire de Paris, CNRS, Universite Paris Diderot, 5 Place Jules Janssen, 92190 Meudon, France\\
             }

   \date{Received September ??, 2015; accepted ??? ??, ????}

\abstract
  {The current detectors of gamma-ray emission specially at TeV energies have too poor resolution to determine whether this emission is produced in the jet or in the core, particularly   of low luminous, non-blazar AGNs (like radio galaxies). In recent works it has been found that the power released by events of turbulent fast magnetic reconnection in the core region of these sources  is more than sufficient to reproduce the observed gamma-ray luminosities. Besides,  3D MHD simulations with test particles  have demonstrated that a first-order Fermi process within  reconnection sites with embedded turbulence results very efficient particle acceleration rates.} 
  {We computed here the spectral energy distribution (SED) from radio to gamma-rays of  the radio galaxies for which energy emission up to TeVs has been detected  (namely, M87, Cen A, Per A, and IC 310).} 
   {For this aim, we employed the acceleration model  above and considered all  the relevant leptonic and hadronic loss processes around the core region of the sources.}
{We found that the calculated SEDs  match very well specially with the VHE observations, 
 therefore strengthening the conclusions above in favour of a core emission origin for the VHE emission of these sources. The model also naturally  explains  the observed very fast variability of the VHE emission. }
  
\keywords{Magnetic reconnection --
                particle acceleration --
                radiation mechanisms: non-thermal
               }

   \maketitle

\section{Introduction}

The non-thermal multi wavelength emission from active galactic nuclei (AGNs) has been broadly studied. Regarding  the very high energy (VHE) emission, until recently only AGNs with highly beamed jets towards the line of sight, namely  blazars,   were detected by gamma-ray telescopes.  More than a  chance coincidence, these detections  are consistent with the conventional scenario that attributes the VHE emission of these sources to particle acceleration along the jet being strongly Doppler boosted and producing  apparently very high fluxes.

Lately, however, a few non-blazar sources which belong to the branch of low luminosity AGNs (or simply LLAGNs)  for having   bolometric luminosities of only a few times the Eddington luminosity, $L_{Edd}$ (\citealt{ho_etal_97, nagar_etal_05})
have been also detected at TeV energies by  ground based $\gamma$-ray observatories (e.g., \citealt{sol2013} and references therein). 
The angular resolution and  sensitivity of these detectors are still  so  poor that it is hard to   establish exactly the location of the emission, i.e. whether it comes from the jet or the core (e.g., \citealt{kachel10}).

Among these sources, the radio galaxies M$87$, Centaurus A (Cen A or NGC $5128$), Persus A (Per A or NGC $1275$) and IC 310 are probably the most striking cases. These VHE detections were surprising because, besides being highly underluminous, the viewing angle of the jets of these sources is of several degrees, therefore allowing for only moderate Doppler boosting. 
These  characteristics make it difficult explaining the VHE of these sources adopting  the same standard scenario of blazars.

Furthermore, observations by \textit{MAGIC}, \textit{HESS} and \textit{VERITAS} 
of  short time scale variability in the $\gamma$-ray emission  of  IC 310, M87 and Per A (\citealt{ahar06,abdo09c,ackermann12,aleksic10a,aleksic10b,aleksic12a,aleksic12b, aleksic14c,aleksic14b}) indicate that it  is produced in  a very compact region that might be the core. 
 In the case of Cen A, though  there is no evidence of significant variability  at $E>$100\rm{MeV},  \rm{GeV} or \rm{TeV} bands  by \textit{Fermi-LAT}(\citealt{abdo10}) or  \textit{HESS} (\citealt{ahar09}), it has been also argued that   the $HESS$ data of this source would be more compatible with a point source near the core  (\citealt{kachel10}). If the $\gamma-ray$ photons were due, for instance,  to proton-proton ($pp$) interactions along the jet then on leaving the source they would interact with the extragalactic background light (EBL) resulting in a flatter spectrum in the TeV range which is not  compatible with \textit{HESS} measurements (\citealt{kachel09b}, see however, other potential explanations in \S. 5).

Though a number of works have attempted to explain these observations as produced in the large scale jets of these sources (e.g., \citealt{stawarz06}),  the evidences above led to the search for alternative particle acceleration scenarios involving the production of the VHE in the surrounds of the nuclear black hole (BH), for instance, in a pulsar-like cascade mechanism in the BH magnetosphere (e.g., \citealt{neronov07}), or in the  sub-parsec scale jet (e.g., \citealt{tavecchio08,abdo09a}). In particular, \cite{tavecchio08}, invoked a two zone model
with  a jet with a fast spine and a slower layer to explain the TeV flares, while \cite{lenain08} proposed that the emission would occur while the jet is collimating, and  \cite{georga05} while it is  decelerating. Another process to  explain these VHE flares   has been proposed by \cite{gian10} in which misaligned mini-jets driven by magnetic reconnection  are moving within the jet with  relativistic velocities relative to it. A two-step acceleration model to TeV energies was also proposed by \cite{istomin09} in the surrounds of the BH involving initial particle acceleration within the accretion disk and then further centrifugal acceleration in the rotating magnetosphere.

In this work, we consider an alternative model in which particles are accelerated, through  a first-order Fermi process, in the surrounds of the BH by the magnetic  power extracted from fast magnetic reconnection events occurring between the magnetosphere of the BH  and the magnetic field lines arising from the inner accretion disk (see Figure~\ref{fig1}). Inspired by similar phenomena occurring  in  space environments, like the earth magnetotail and the solar corona, 
\cite{gl05}    explored this process first in the framework of microquasars and then  \cite{beteluis2010a}  and \cite{beteluis2010b} extended it  to  AGNs. 

In these works,  \cite{gl05} and \cite{beteluis2010a} found that fast reconnection could be efficient enough to produce the core radio outbursts in microquasars and AGNs. More recently,  \cite{beteluis2014} (henceforth KGS15)  and \cite{chandra14} revisited this model exploring different mechanisms of fast magnetic reconnection and accretion,  and extending the study to include also the gamma-ray  emission of a much larger sample containing more than 230 sources. They confirmed the earlier trend found by \cite{gl05} and \cite{beteluis2010a}, and verified that if fast reconnection is driven by turbulence (\citealt{LV99}) 
there is a correlation between the fast magnetic reconnection power and the  BH mass spanning $10^{10}$ orders of magnitude  that can explain not only the observed radio, but also the  VHE  luminosity from  microquasars and LLAGNs (involving all the sources of 
the  so called fundamental plane of BH activity \citealt{merloni}).

The correlations found  in the works above (specially in KGS15 and \citealt{chandra14})                                       between the calculated power released by magnetic reconnection in the surrounds of the BH and the observed radio and gamma-ray luminosities of a very large sample of LLAGNs and microquasars, have  motivated further investigation to test the validity of the model. 
In particular, in a recent work, we explored in detail the non-thermal  emission of the microquasars Cygnus X-1 and Cygnus X-3 and found that this reconnection acceleration model 
is able to reproduce most of the features of their observed spectral energy distribution (SED)  in outburst states up to TeV energies (\citealt{khiali14} hereafter KGV15).

Our aim here is to extend this  study to the supermassive BH sources of the KGS15 and \cite{chandra14} sample, trying   to reconstruct the observed SEDs, specially at the VHE branch, of the four   radio galaxies mentioned earlier, i.e.,  Centaurus A, Per A, M87 and IC 310 
applying the same acceleration model  above.   

We first compute the power released by fast magnetic reconnection in the surrounds of the BH as described, and then the resulting particle spectrum of the accelerated particles in the magnetic reconnection site. In particular, we explore the first-order Fermi  acceleration process that may occur within the current sheet as  proposed in \cite{gl05}. Such acceleration mechanism has been extensively 
tested numerically 
in 3D collisional MHD simulations of magnetic current sheets employing test particles
 (\citealt{kowal2011,kowal2012,bete14}) and also  in collisionless particle in cell simulations (e.g.,  \citealt{drake06,zenitani09,drake10,cerutti13,cerutti14}; see also the reviews by \citealt{bete14,betegreg15}).

In order to reconstruct the SED, we consider the relevant radiative processes due to the interactions of the accelerated particles  by magnetic reconnection with the surrounding radiation, matter and magnetic fields. We then compare the rates of these radiative losses with the magnetic reconnection acceleration rate and  determine the maximum energy that the electrons and protons can attain.  
For comparison, we also consider the acceleration rate due to shocks in the surrounds of the reconnection region, but find that this is less effective than the acceleration by magnetic reconnection.

We show for the first time that a consistent and numerically tested acceleration model by magnetic reconnection in the surrounds of BH sources can effectively reproduce the observed SEDs of the four radio galaxies up to TeVs.

The outline of the paper is as follows. In Section 2, we describe in detail our scenario presenting both the acceleration model and the emission mechanisms. In section 3, we show  the results of the application of the acceleration and emission model to Cen A, Per A, M87 and IC 310. Finally,  we discuss and summarize our results  drawing our conclusions in Section 4.

\section{Description of the acceleration and emission model}

\begin{figure}
 \centering

 \includegraphics[width=3.5in]{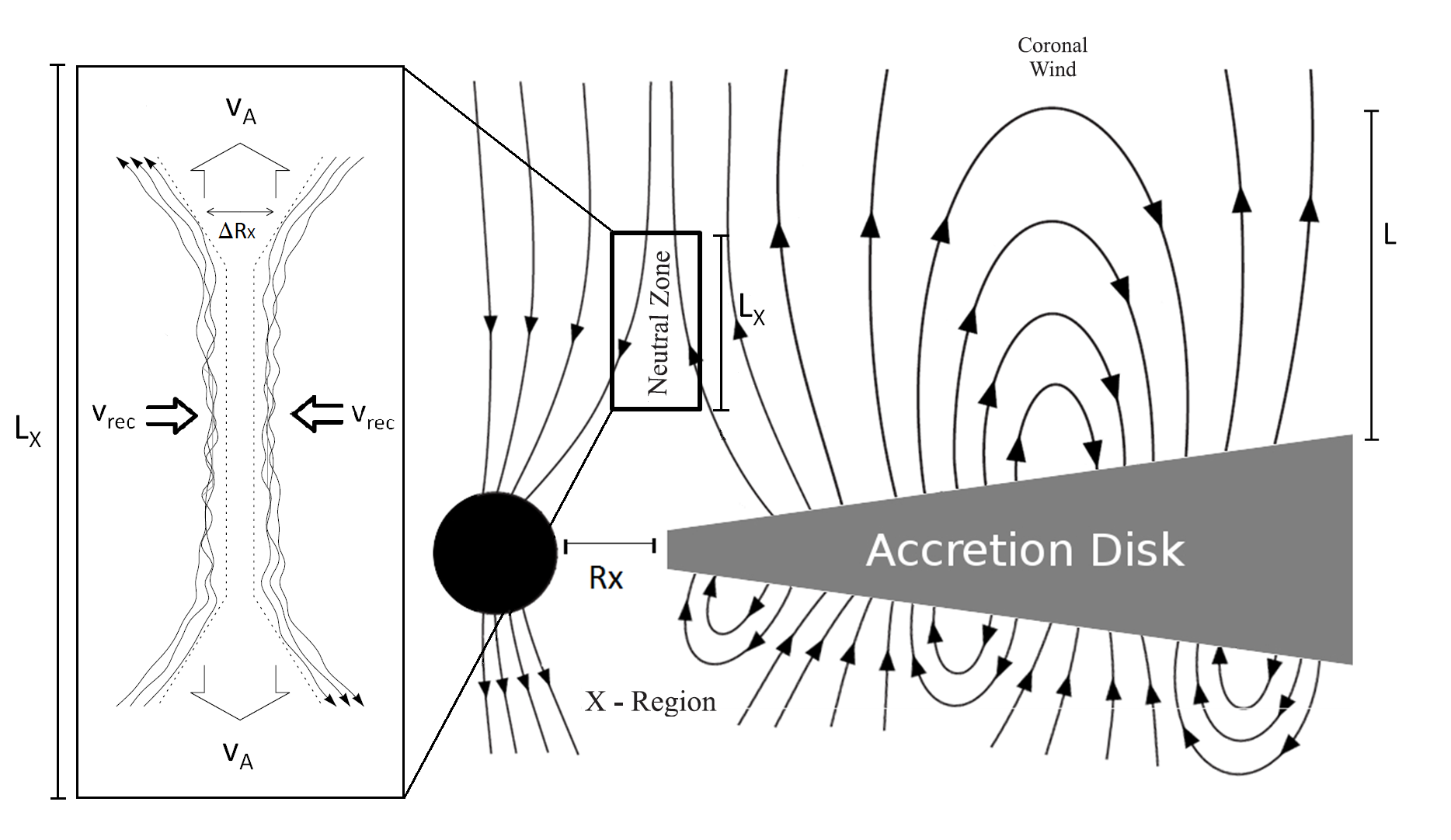}
 \caption{Scheme of magnetic reconnection between the lines rising from the accretion disk into the corona and the lines around the BH horizon. Reconnection is made fast by the presence of embedded turbulence in the reconnection zone (as indicated in the detail).  Particle acceleration may occur in the magnetic reconnection zone by a first-order Fermi process (adapted from \citealt{gl05}).}
 \label{fig1}
\end{figure}

As stressed, we consider here that  relativistic particles may be accelerated in the core of  LLAGNs,  i.e., in the coronal region around the BH near the basis of the jet launching, as a result of events of  fast magnetic reconnection and examine whether this process may reproduce the observed emission specially at VHEs.  This  model has been described in detail by \cite{gl05} and \cite{beteluis2010a}, and more recently by KGS15. We summarize below its main characteristics. 
We assume that the inner region of the accretion disk/corona system  alternates between two states which are controlled by changes in the global magnetic field. 
Right before a fast magnetic reconnection event, we adopt the simplest possible configuration  
by considering a magnetized standard geometrically thin and optically thick accretion disk around the BH as in the sketch of Figure~\ref{fig1} (see also \citealt{chandra14} for an alternative solution considering a magnetic-ADAF accretion disk).

The magnetosphere around the central BH can  be built from the drag of magnetic field lines by the accretion disk (e.g., \citealt{macdonald86,koide02}).  The  large-scale poloidal magnetic field in the disk corona can in turn be formed by the action of a  dynamo inside the accretion disk 
or dragged from the surroundings.
Under  the action of  disk differential rotation, this poloidal magnetic flux gives rise to a wind that partially removes angular momentum from the system and increases the accretion rate.
This, in turn, increases the ram pressure of the accreting material that will press the magnetic lines in the inner disk region against the lines of the BH magnetosphere thus favouring the occurrence of  reconnection (see Figure~\ref{fig1}). 
We note that according to mean field dynamo theory, an inversion of the polarization of the  magnetic lines is expected to occur every half of the dynamo cycle; when this happens a new flux of disk lines should reach the inner region with an inverted polarity with respect to the magnetic flux already sitting around the BH, therefore, favouring magnetic reconnection between the two fluxes. The advection of field lines from the outer regions  also allows for periodic changes in the polarity. Also for simplicity, we assume the co-rotation between the inner disk region and the BH magnetosphere. 

The magnetic field intensity in this inner region can be determined from the balance between  the magnetic pressure of the BH magnetosphere and the accretion ram pressure and is given by (KGL15):

\begin{equation} \label{1}
B\cong 9.96\times 10^{8}r_X^{-1.25} \xi^{0.5}m^{-0.5}\ {\rm G},
\end{equation}
where $m$ is the BH mass in units of solar mass, $\xi$ is the  mass accretion disk rate in units of the Eddington rate ($\xi=\dot{M}/\dot{M}_{Edd}$, with  $\dot{M}_{Edd} =1.45\times10^{18}  m$ g s$^{-1}$), and  $r_X=R_X/R_S$ is the inner radius of the accretion disk in units of the BH Schwarzchild radius ($R_S=\frac{2GM}{c^2}$).  As  KGS15,  we adopt here $r_X=6$.

\subsection{Conditions for fast reconnection in the surrounds of the BH}

As discussed in KGS15,  the presence of embedded turbulence in the nearly collisional MHD coronal flow of the core region of the AGNs can make reconnection   very  fast  (e.g., \citealt{LV99}) and  cause the release of large amounts of magnetic energy power in the scenario described in Figure~\ref{fig1}.
\footnote{We note that the strongly magnetized and low dense coronal fluid of the systems we are considering  in this work satisfies the condition  $L> l_{mfp} > r_l$ (where $L$ is a typical large scale dimension of the system, $l_{mfp}$ the ion mean free path and $r_l$ the ion Larmor radius). For such flows a weakly collisional or effectively collisional MHD description is more than appropriate  and  we will employ this approach here (see more details in KGS15).}

According to the model proposed by \cite{LV99}, even weak embedded turbulence causes the wandering of the magnetic field lines which allows for many independent patches to reconnect simultaneously making the global reconnection rate  large, $V_R \sim v_A (l_{inj}/L)^{1/2} (v_{turb}/v_A)^2$, where $V_R$ is the reconnection speed, $v_A$ is the Alfv\'en speed, and $l_{inj}$ and  $v_{turb}$ the injection scale and velocity of the turbulence, respectively. This expression indicates that the reconnection rate can be as large as $\sim V_A$.

This theory has been deeply  investigated (e.g.  \citealt{eyink11,lazarian12}) and  confirmed numerically 
(\citealt{kowal09, kowal2012}). In particular, it has been shown  (\citealt{eyink11}) that turbulent collisional fast reconnection prevails when the thickness of the magnetic discontinuity layer  (see Eq.~\ref{3} below, and Figure~\ref{fig1})  is larger than the ion Larmor radius. As demonstrated in KGS15,  for the systems we are studying this condition is naturally satisfied and we will adopt this process to derive the magnetic power released by fast reconnection.
We should also notice that there has been direct evidences of turbulent reconnection in solar coronal  events   
(e.g., \citealt{priest01}) and also in the Earth magnetotail (\citealt{retino07}).

The fluids we are investigating here have  large hydrodynamical and magnetic Reynolds numbers (KGS15) 
implying  that they can be easily  distorted and become turbulent. For instance,  
  current driven instabilities, can naturally drive turbulence with characteristic velocities around the particles thermal speed.  Also, the occurrence of continuous slow magnetic reconnection during the building of the corona itself in the surrounds of the BH \citep{liu03} will contribute to the onset of turbulence which will then be further fed by fast reconnection as in the \cite{LV99} model (see \citealt{oishi15,lazarian15}). Numerical simulations of coronal disk accretion also indicate the formation of turbulent flow in the surrounds of the BH 
\citep[see e.g,][]{dexter_etal_14}. All these processes may ensure the presence of embedded weak turbulence in the magnetic discontinuity described in Figure~\ref{fig1}.
\footnote{We note that in the  model described here,  the turbulence  is in general sub-Alfv\'enic due to the strong magnetic fields implied  and nearly  tran-sonic (since the turbulent velocity is of the order of the sound speed and smaller than the Alfv\'en speed)  and therefore, incompressible.  This regime of turbulence has been extensively investigated in the literature (see e.g., \citealt{LV99,lazarian12}) and  the acceleration formulae employed here have been obtained directly from numerical MHD simulations with particle tests injected in current sheets with embedded turbulence also implying this regime (Kowal, de Gouveia Dal Pino, Lazarian 2012).}

The  magnetic reconnection power released   by turbulent driven fast reconnection in the magnetic discontinuity region (as schemed in Figure~\ref{fig1}), has been derived in KGS15 and is given by:
\begin{equation} \label{2}
W \simeq 1.66\times 10^{35} \psi^{-0.5} r_X^{-0.62} l^{-0.25} l_X q^{-2}\xi^{0.75}m\ ~ {\rm erg~s^{-1}},
\end{equation}
where  $l=L/{R_S}$ is the height of the corona in units of $R_S$; $l_X={L_X}/{R_S}$, $L_X \leq L$ is the extension of the magnetic reconnection zone (as shown in Figure~\ref{fig1}; see also Table 1), $q=[1-(3/r_X)^{0.5}]^{0.25}$ and $v_A = v_{A0} \psi$ is the relativistic form of the Alfv\'en velocity,   with $v_{A0} = B/(4\pi \rho)^{1/2}$,  $B$ being the local magnetic field,  $\rho \simeq n_c m_p$  the fluid density  in the corona, $n_c$ the coronal  number density, $m_p$ the proton mass,  and  $\psi=[1+({v_{A0} \over c})^2]^{-1/2}$,  
in this  work,  $v_{A0} \sim c$.
 The results of KGS15 (Figure 5 in KGS15) have shown that accretion rates $\xi$  between $0.05 < \xi \leq  1$  are able to produce magnetic reconnection power values which are larger than the observed luminosities of LLAGNs. We  adopt here $\xi \simeq 0.7$, but we should notice that the results are not very much sensitive to the choice of this parameter.  As demonstrated in KGS15 and SKG15 studies, one can match the observations by taking alternative fiducial combinations of the free parameters in the equation above, particularly by constraining the size of the height of the corona, $L$.

We will employ the equations above  in Section 3 to model the acceleration of particles in the core of Cen A, Per A, M87 and IC 310.
The acceleration region in our model corresponds to the cylindrical shell around the BH where magnetic reconnection takes place (see Figure~\ref{fig1}). This shell has a length $l_X$, with  inner and outer radii  given by $R_X$ and $R_X+\Delta R_X$ respectively, where $\Delta R_X$ is the width of the current sheet  given by (KGS15): 
\begin{equation} \label{3}
\Delta R_X\cong 2.34\times 10^{4} \psi^{-0.31} r_X^{0.48} l^{-0.15} l_X q^{-0.75}\xi^{-0.15}m\ {\rm cm}.
\end{equation}

In \S. 3, we will also need the accretion disk temperature in order to evaluate the black body radiation field:

\begin{equation} \label{4}
T_d\cong 3.71\times 10^{7}\alpha^{-0.25}r_X^{-0.37}m^{0.25}\ \rm K,
\end{equation}
where $0.05 \leq \alpha< 1$ is the Shakura-Sunyaev disk viscosity parameter which we here assume to be of the order of 0.5. 

In addition, we will need the coronal number density to compute the radiative losses of accelerated particles which is (KGS15)
\begin{equation} \label{11}
n_c\cong 8.02\times 10^{18}r_X^{-0.375} \psi^{0.5} l^{-0.75}q^{-2}\xi^{0.25}m^{-1}\ {\rm cm^{-3}},
\end{equation}
while the coronal temperature is given by
\begin{equation} \label{11b}
T_c\cong 2.73\times 10^{9}r_X^{-0.187} \psi^{0.25} l^{0.125}q^{-1}\xi^{0.125}\ {\rm K}.
\end{equation}

The magnetic power in Eq.~\ref{2} heats the surrounding gas and accelerates particles. 
As in KGV15, we assume that approximately $50\%$ of the reconnection power is used to accelerate the particles. This  is consistent with  plasma laboratory experiments of particle acceleration in reconnection sheets (e.g., \citealt{yamada}) and also with the observations of flares in the Sun  (e.g., \citealt{lin71}).

\subsection{Particle acceleration due to magnetic energy released by fast reconnection}
 
A first-order Fermi acceleration may occur when particles of the fluid are trapped between the two converging magnetic flux tubes moving to each other  in the magnetic reconnection discontinuity with a velocity $V_{R}$. \cite{gl05} first investigated this process analytically and  showed that, as the particles bounce back and forth undergoing head-on collisions with magnetic fluctuations in the current sheet, their energy increases by $< \Delta E/E > \sim 8 V_{R}/3c$ after each round trip, which leads to an exponential energy growth after several round trips. 
If  magnetic reconnection is fast,  $V_{R}$ is of the order of the local Alfv\'en speed $V_{A}$ and,  at the surroundings of relativistic sources $V_{R} \simeq  v_A \simeq c$ and thus  the mechanism can be rather efficient.

From the results of  3D  MHD numerical simulations of this  process \citep{kowal2012}, we find that the  acceleration rate for a proton is given by (KGV15):
\begin{equation} \label{5}
t^{-1}_{acc,rec,p}=1.3\times 10^5\left(\frac{E}{E_0}\right)^{-0.4}t_0^{-1},
\end{equation}
where $E$ is the energy of the accelerated proton, $E_0=m_p c^2$, $m_p$ is the proton rest mass, $t_0=l_{acc}/{v_A}$  is the Alfv\'en time,  and $l_{acc}$ is the length scale of the acceleration region.

Similarly,  for the electrons one can get:
\begin{equation} \label{6}
t^{-1}_{acc,rec,e}=1.3\times 10^5\sqrt{\frac{m_p}{m_e}}\left(\frac{E}{E_0}\right)^{-0.4}t_0^{-1},
\end{equation}
where $m_e$ is the electron rest mass.

The two equations above do not take into account the effects of radiative losses upon the accelerated particles. They will be used to compute the acceleration rates in our model.

As stressed in \cite{gl05}, it is also possible that a diffusive shock may develop in the surrounds of the magnetic reconnection zone at the jet launching region caused by plasmons or coronal mass ejections. As in  solar flares,  these can be produced in the  reconnection layer and released along the  magnetic field lines. In this case, we expect the shock velocity to be predominantly parallel to the magnetic lines. and the  acceleration rate  for a particle of energy $E$  will be approximately given by  (e.g.,  \citealt{spruit88}):
 \begin{equation} \label{7}
 t^{-1}_{acc,shock}=\frac{\eta e c B}{E},
 \end{equation}
where $0<\eta \ll 1$ characterizes the efficiency of the acceleration. We fix $\eta=10^{-1}$, which is appropriate for shocks with velocity $v_s\approx 0.1c$  commonly assumed in the Bohm regime.

In \S. 3 we  compare both the magnetic reconnection acceleration time (Eqs.~\ref{5} \& \ref{6}) and  the shock acceleration time  (Eq.~\ref{7})  with the relevant radiative cooling process that cause the loss of energy of the accelerated particles and constrain their maximum energy. 
These particles may lose  energy 
via interactions with the surrounding magnetic field (producing synchrotron emission), the photon field (producing inverse Compton, synchrotron-self-Compton, and photo-meson interactions), and with the surrounding matter (producing pp collisions and relativistic Bremsstrahlung radiation).
In \S. 2.4,  we discuss the relevant radiative loss processes for electrons and protons which will allow the construction of the SED of the  sources  M87, Cen A, Per A and IC 310 for comparison with the observations.

\subsection{Particle energy distribution}

The accelerated particles are expected to develop a power law spectrum. Their injection and cooling will occur mainly in the coronal region around the black hole (see Figure~\ref{fig1}).
The isotropic injection function (in units of $\rm{erg^{-1} {\rm cm^{-3} s^{-1}}}$) is given by (see e.g. KGV15):
\begin{equation} \label{8}
Q(E)=Q_0 E^{-p}exp{[-E/E_{max}]}
\end{equation}
with $p>0$ and $E_{max}$  the cut-off energy which can be calculated by the balance of acceleration and the energy losses. Particles can gain energy up to a certain value $E_{max}$ for which the total cooling rate equals the acceleration rate. 

We assume for the power law index $p$ values between 1 and 2.5 for the sources here investigated (see \S. 3), which are compatible with  the analytical and numerical studies of particle acceleration within magnetic reconnection sheets
(e.g., \citealt{Drury2012,kowal2012}).

The normalization constant $Q_0$ is calculated from the total power injected in each type of particle
\begin{equation} \label{9}
L_{(e,p)}=\int_{V} d^3r \int_{E_{min}}^{E_{max}}dE\ E\ Q_{(e,p)}(E)
\end{equation}
where $V$ is the emission volume in the surrounds of the magnetic reconnection  acceleration region in Figure~\ref{fig1} (see \S. 2.5) and $L_{(e,p)}$  is the fraction of the magnetic reconnection power that accelerates the electrons and protons calculated from Eq.~\ref{2}. This injected power is equally shared between protons and electrons.

The kinetic equation that describes the general evolution of the particle energy distribution $N(E, t)$ is the Fokker-Planck differential equation \citep{ginzburg64}.  We here use a simplified form of this equation. 
We employ the one-zone approximation to find the particle distribution, assuming that the acceleration region is spatially thin enough, so that we can ignore spatial derivatives in the transport equation. Physically, this means that we are neglecting the contributions to $N(E)$ coming from other regions than  the magnetic reconnection region in the inner accretion disk/corona zone in the surrounds of the BH.
We consider a steady-state particle distribution which can be obtained by setting $\frac{\partial N}{\partial t}=0$ in the Fokker-Planck differential equation, so that the particle distribution equation is
\begin{equation} \label{10}
N(E)=|\frac{dE}{dt}|^{-1} \int_E^\infty Q(E)dE.
\end{equation}
Here $-\frac{dE}{dt}\equiv E t_{cool}^{-1}$, 
where $t_{cool}^{-1}$ is the total cooling rate that can be calculated assuming all the cooling mechanisms (we describe them in the following section briefly).

\subsection{Photon Absorption}
We consider two main absorption processes of the photons produced by the accelerated particles in the nuclear region of the sources:  the gamma-ray photon absorption due to $e^-e^+$ pair creation, and the absorption of optical and X-ray photons due to external interstellar  neutral gas and dust (photon-neutral)  absorption.

\subsubsection{Photon-photon ($\gamma\gamma$) annihilation}
\begin{figure*} 
    \centering
        \subfigure[Cen A]
    {
        \includegraphics[width=2.4in]{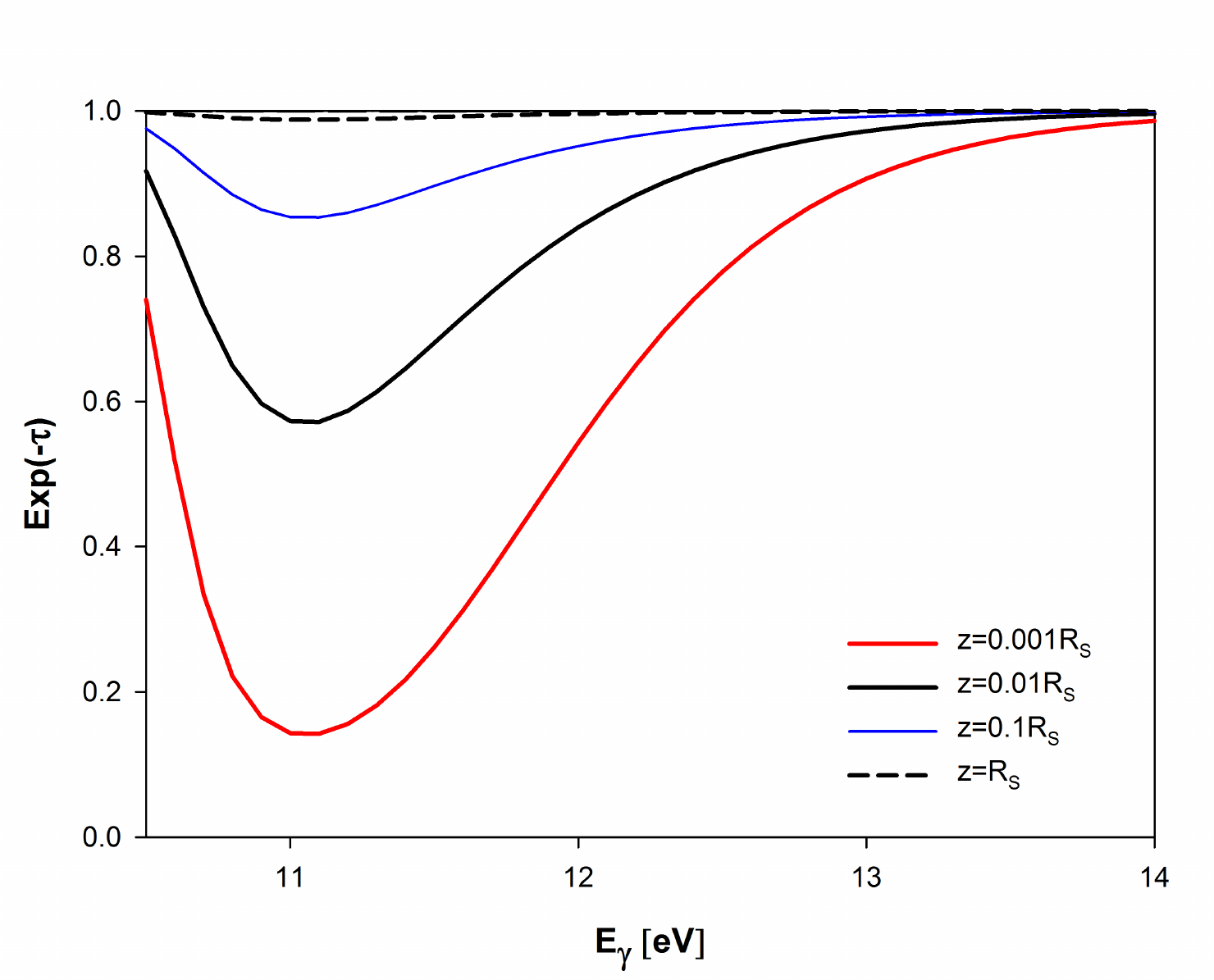}
        \label{8a}
    }
    \subfigure[Cen A]
    {
        \includegraphics[width=2.4in]{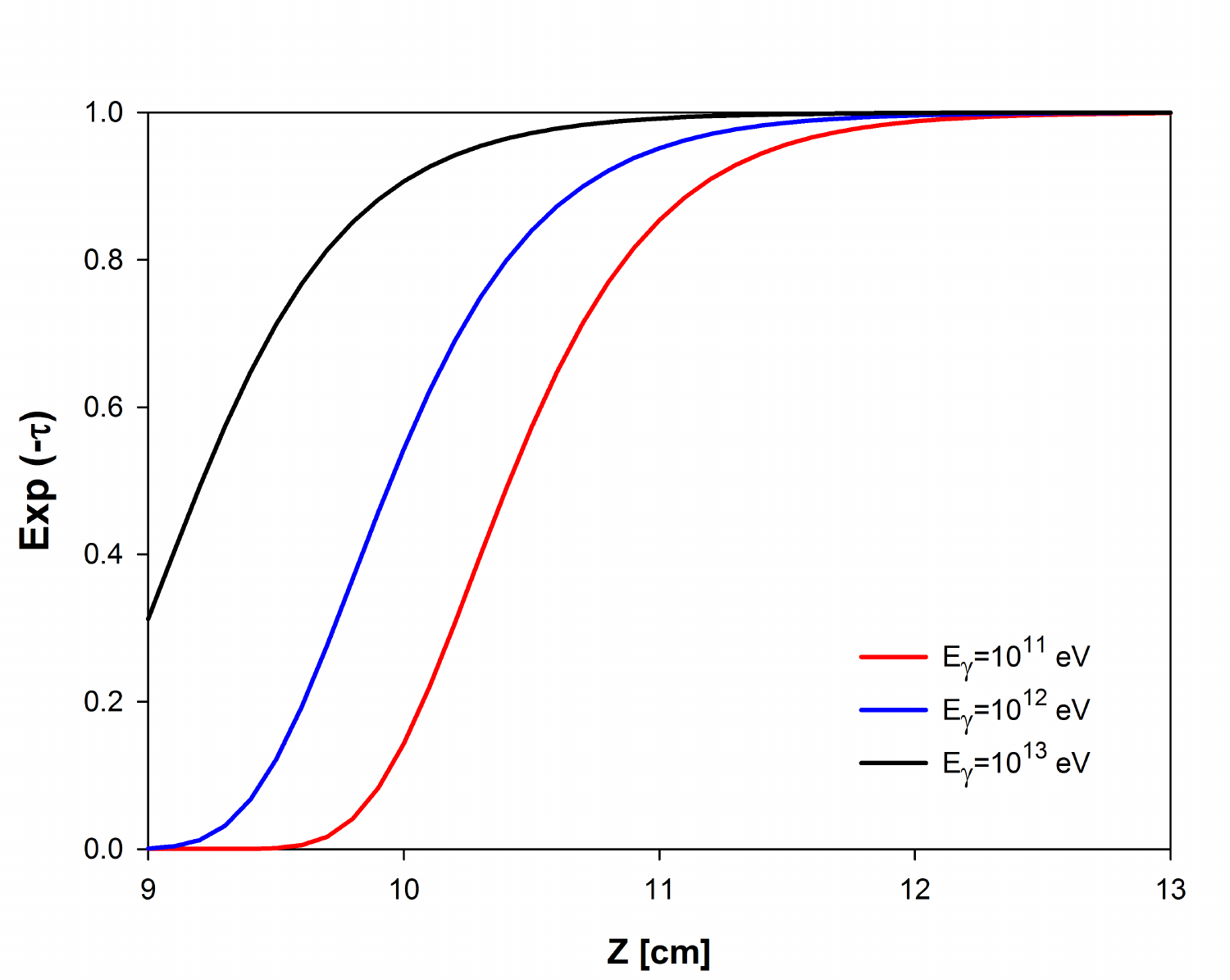}
        \label{8b}
    }
    \\
    \subfigure[Per A]
    {
        \includegraphics[width=2.4in]{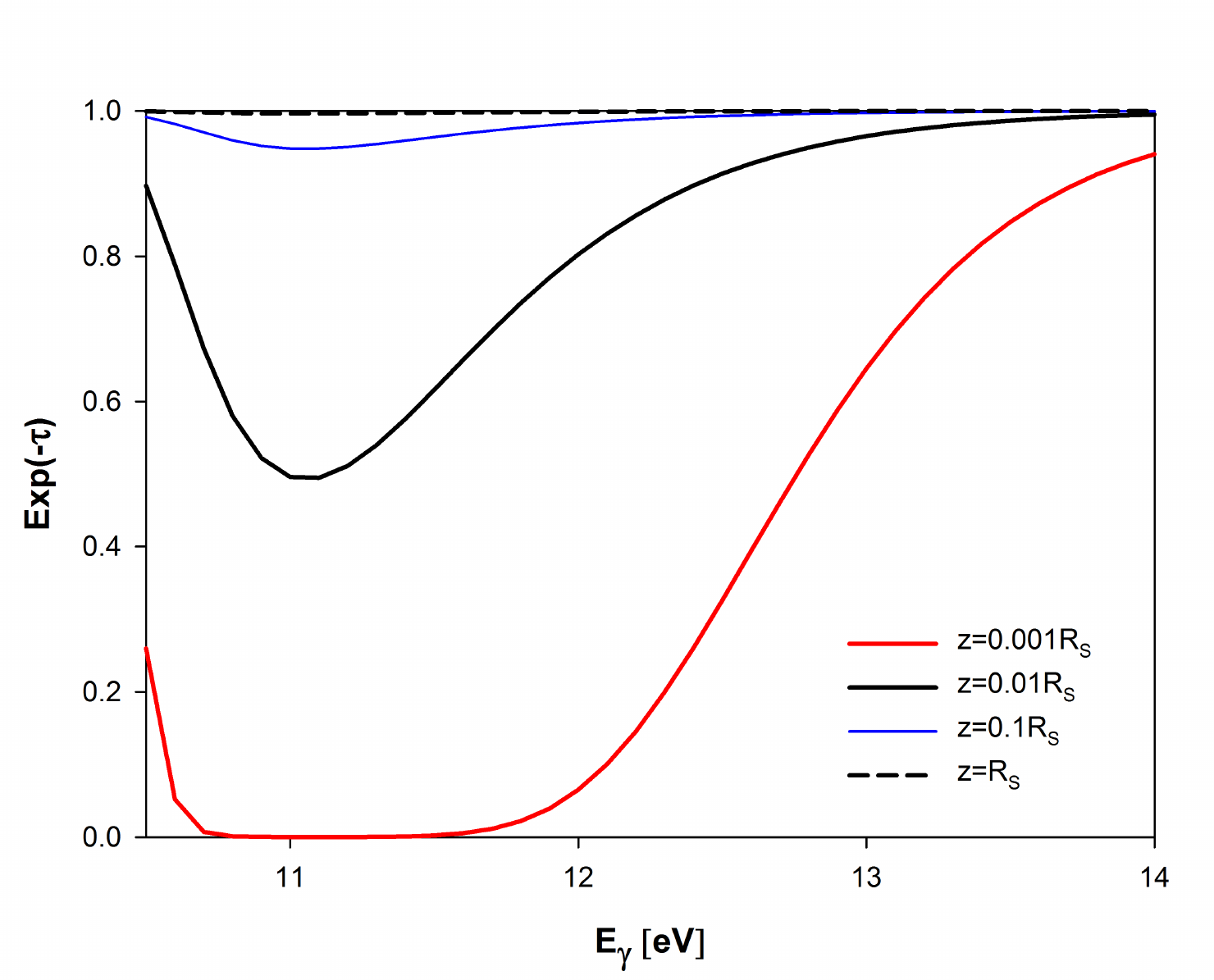}
        \label{8c}
    }
    \subfigure[Per A]
    {
        \includegraphics[width=2.4in]{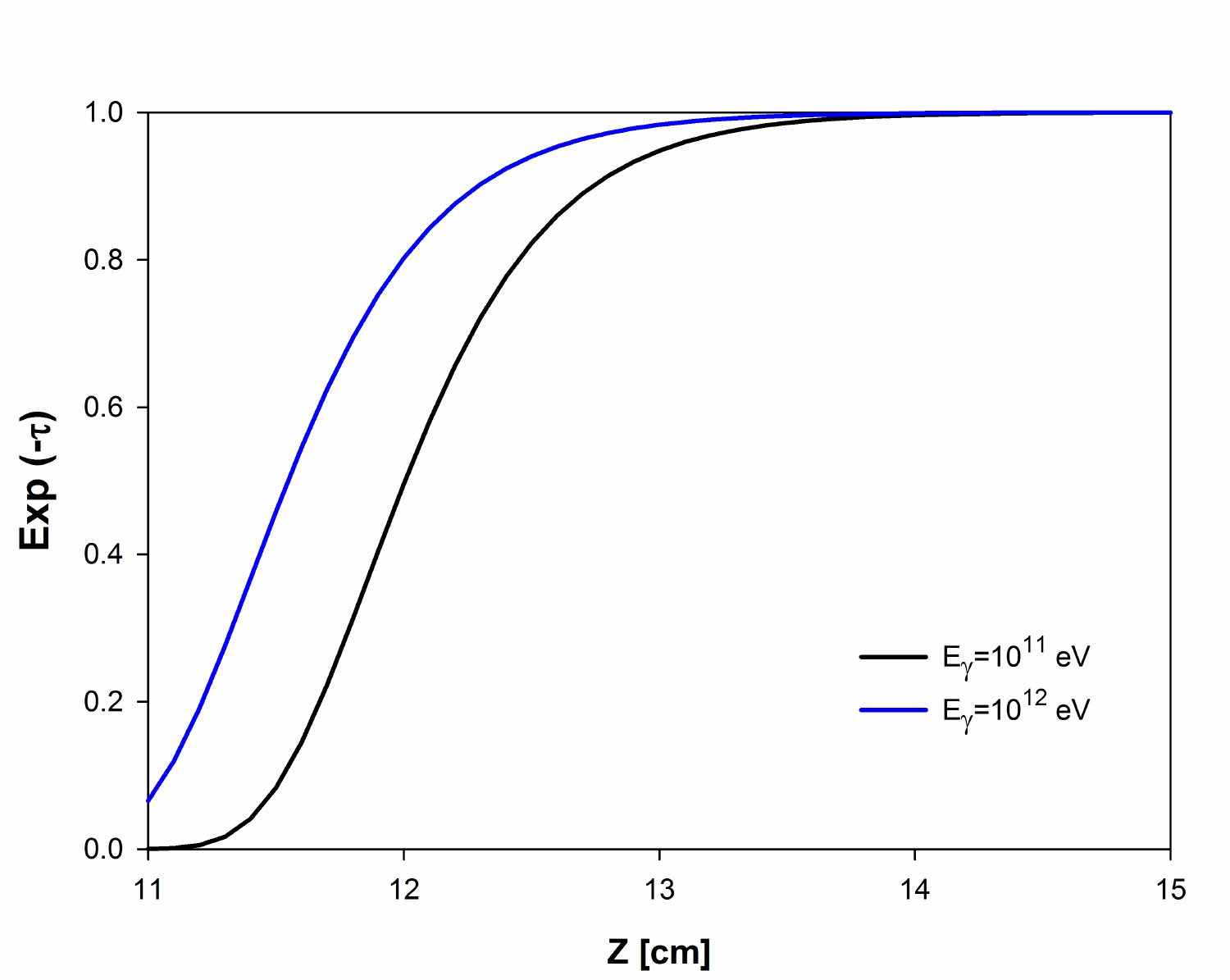}
        \label{8d}
    }
    \\
    \subfigure[M87]
    {
        \includegraphics[width=2.4in]{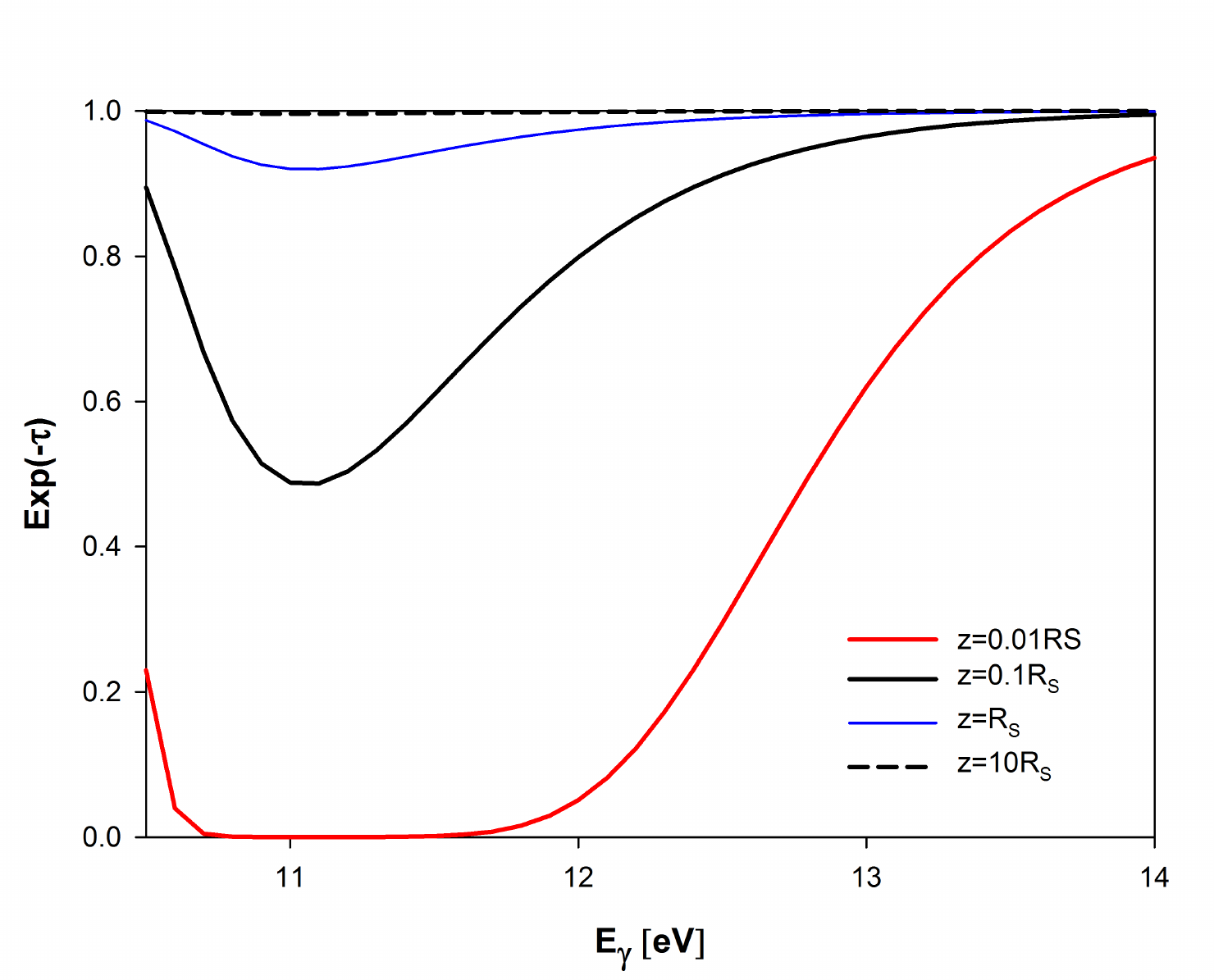}
        \label{8e}
    }
    \subfigure[M87]
    {
        \includegraphics[width=2.4in]{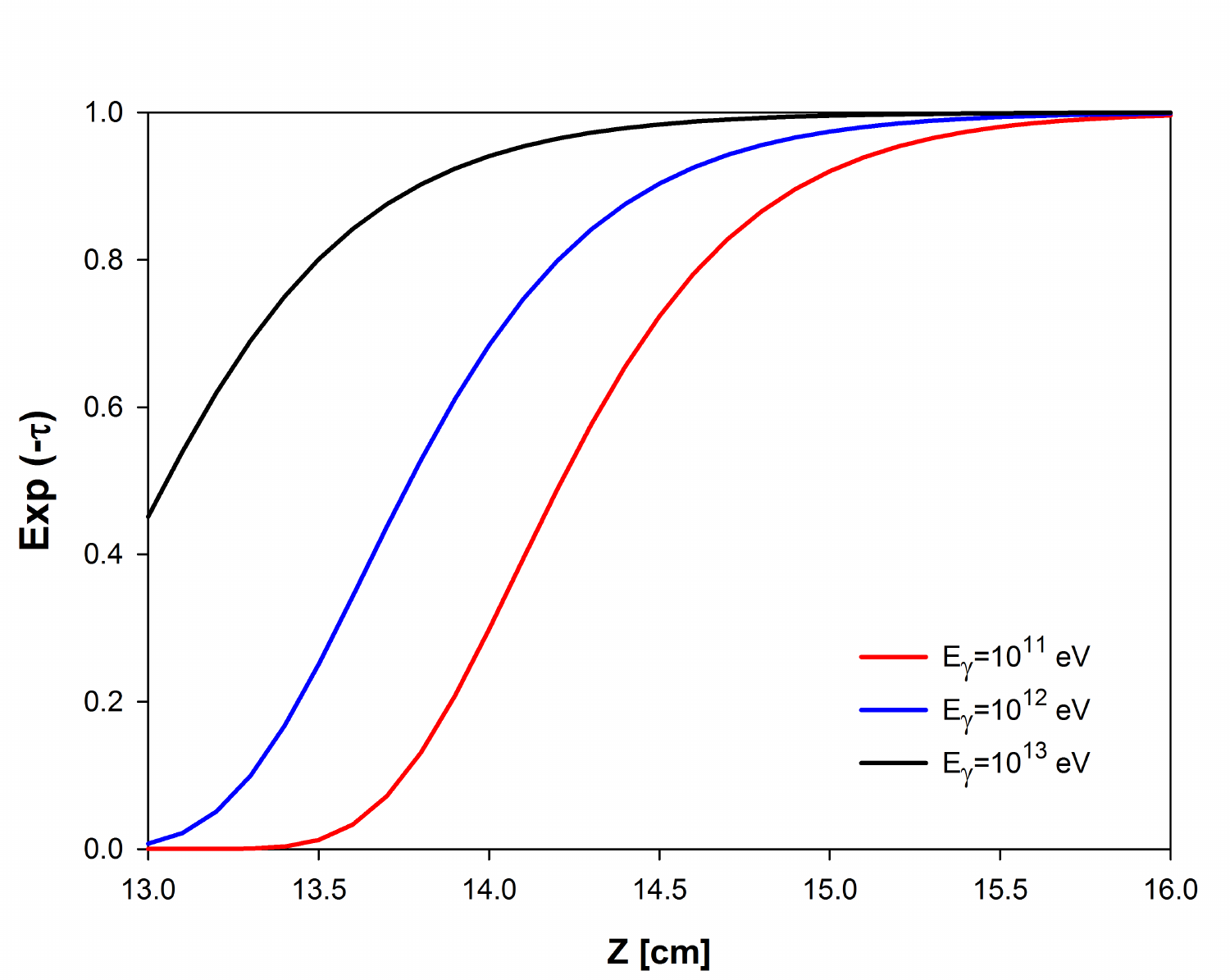}
        \label{8f}
    }
    \\
    \subfigure[IC 310]
    {
        \includegraphics[width=2.4in]{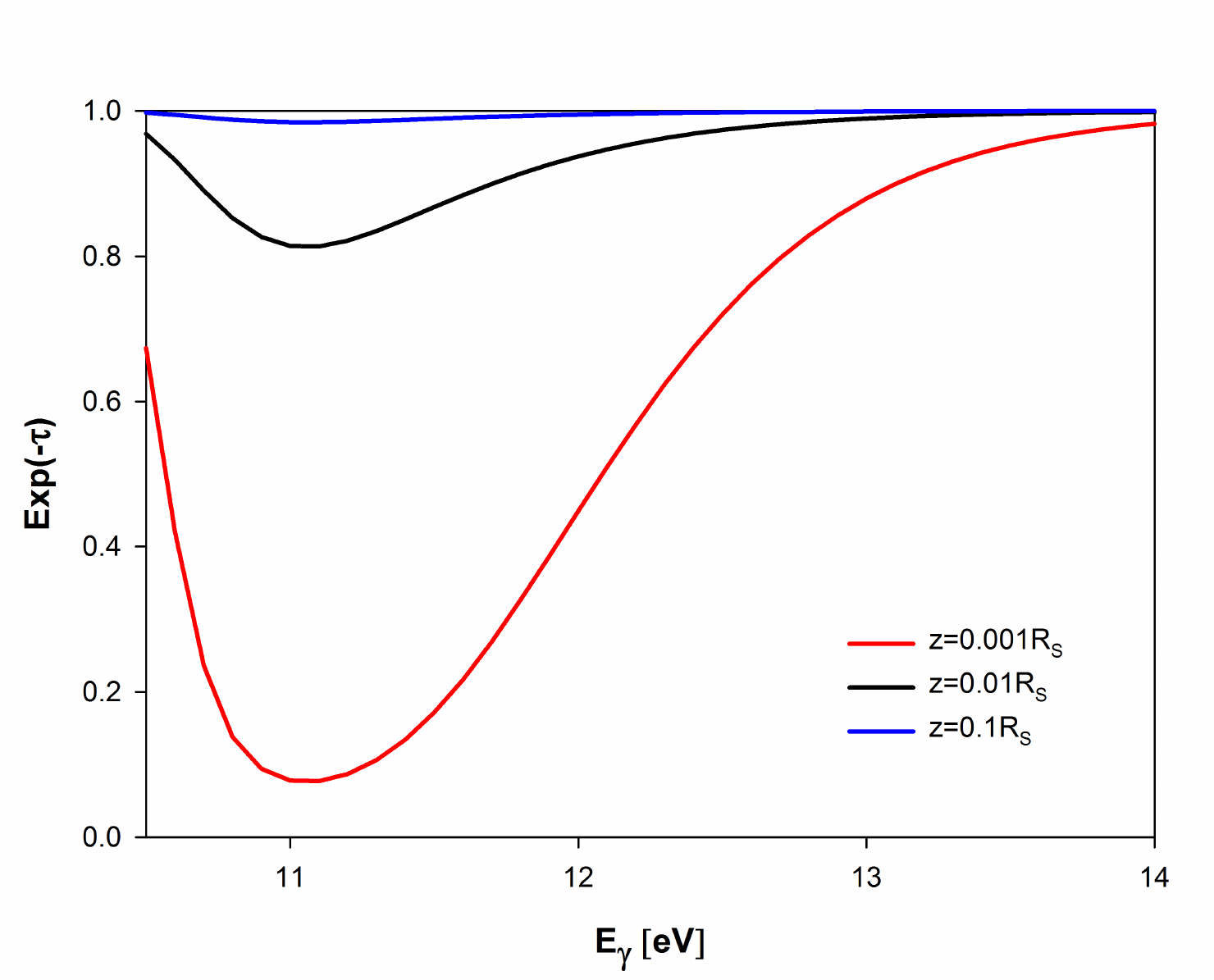}
        \label{8g}
    }
    \subfigure[IC 310]
    {
        \includegraphics[width=2.4in]{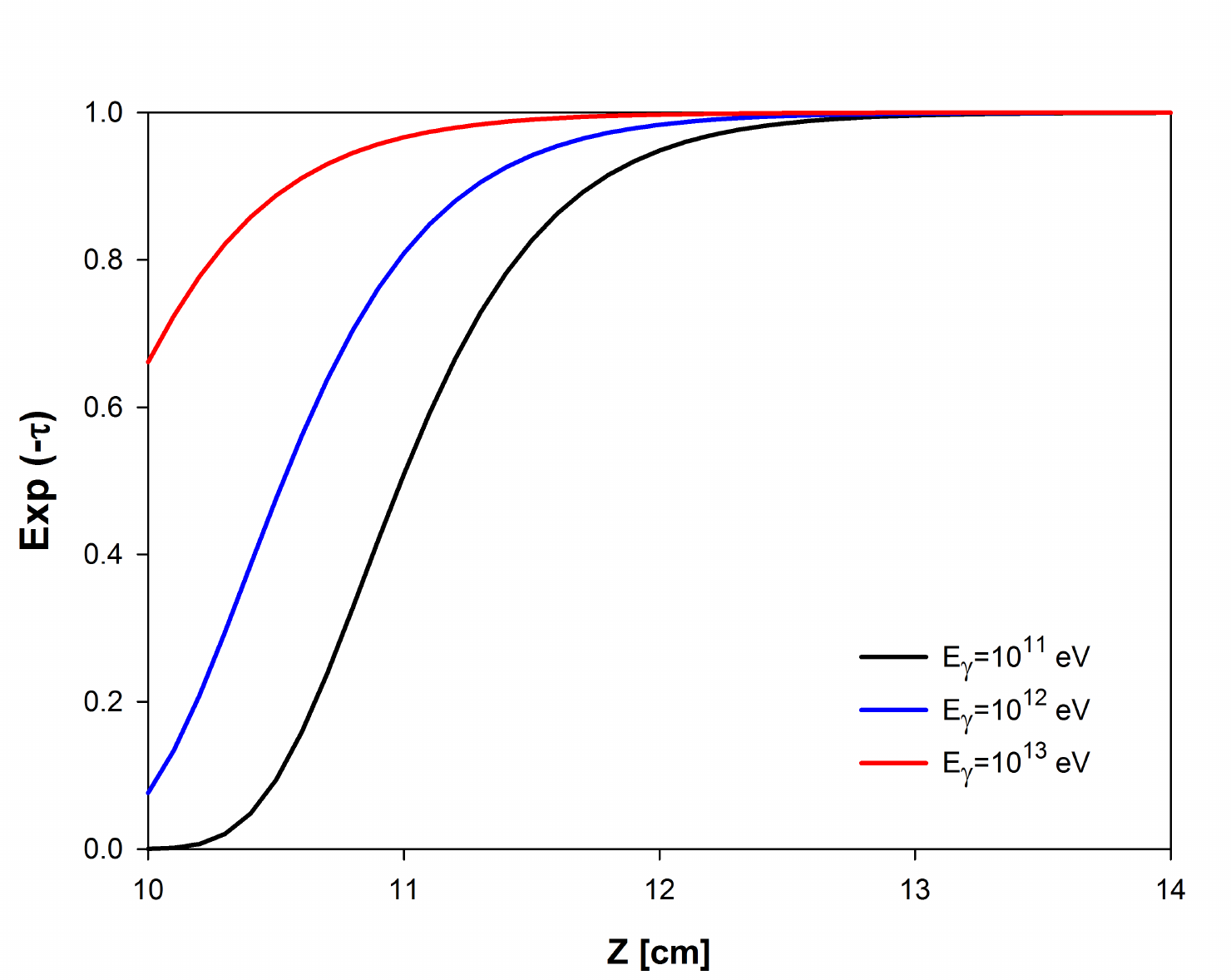}
        \label{8h}
    }
    \caption{Left panels: Spectrum of $\gamma$-ray absorption at selected heights $z$ above the plane of the disk in Cen A, Per A, M87, and IC 310 ($a$, $c$, $e$, and $g$ panels, respectively). Rigt panels: The transmitted flux $exp(-\tau)$ for different $\gamma$-ray energies as function of the height $z$ above the disk in Cen A, Per A, M87, and IC 310 ($b$, $d$, $f$, and $h$ panels, respectively).}
    \label{fig8}
\end{figure*}

The produced  $\gamma$-rays   can be annihilated by the surrounding radiation field via electron-positron pair production, i.e.,  $\gamma+\gamma \to e^++e^-$. In our model the dominant radiation field for this process in the coronal region is due to the scattered photons from the accretion disk (see Figure \ref{fig8}). 
\footnote{We should remark  that we  found  the contribution due to the coronal radiation field itself to be much smaller than that of the  accretion disk. }

 To evaluate the optical depth due to this process, we have adopted the model described in \cite{cerutti11}, assuming that  the $\gamma$-rays are produced within a  spherical region around the disk  with radius extending up to  $L\simeq20R_S$. The attenuated $\gamma-$ray luminosity $L_\gamma(E_\gamma)$  at a  distance $z$ above the disk is given by (\citealt{romero05})
 \begin{equation} \label{12}
L_\gamma(E_\gamma)=L^0_\gamma(E_\gamma)e^{-\tau(z,E_\gamma)}
\end{equation}
 where $L^0_\gamma$ is the intrinsic coronal gamma-ray luminosity and $\tau(z,E_\gamma)$ is the optical depth. The calculated optical depth depends on the $\gamma$-ray energy and the distance above the disk $z$.

Figure~\ref{fig8} depicts the gamma-ray absorption spectrum for the sources investigated here, for different heights z. We see that  in all cases 
 at distances larger  than $\sim0.1R_S$ from the disk surface, the absorption of $\gamma$-rays becomes negligible. Since we are adopting here an emission region with an extension $\simeq 0.3 R_S$ to $\simeq20R_S$,  it is reasonable to exclude the absorption effect above in our calculations of the SEDs (see \S. 3).

\begin{table*}
\centering
\begin{minipage}{160mm}
\caption{\label{llagntable}Model parameters for Cen A, Per A, M87 and IC 310.}
\begin{tabular*}{\textwidth}{@{}llrrrrlrlr@{}}
\hline
 &Parameters &Cen A&Per A&M 87&IC 310\\
\hline
$B$ &Magnetic field (\rm G)&$1.25\times10^4$&$4812$&$1620$&$8874$\\
$W$&Magnetic reconnection power (\rm erg/s)&$1.2\times10^{43}$&$8.2\times10^{43}$&$5.25\times10^{44}$&$4.5\times10^{43}$\\
$\Delta R_X$&Width of the current sheet (\rm cm)&$3.6\times10^{13}$&$2.4\times10^{14}$&$1.35\times10^{15}$&$3.2\times10^{12}$\\
$n_c$&Coronal particle number density ($\rm cm ^{-3}$)&$7.1\times10^{9}$&$10^{9}$&$3.3\times10^{8}$&$10^{11}$\\
$T_d$&Temperature of the disk ($\rm K$)&$1.9\times10^{8}$&$3\times10^{8}$&$5.2\times10^{8}$&$2.25\times10^{8}$\\
$R_x$&Inner radius of disk (\rm cm)&$8.8\times10^{13}$&$6\times10^{14}$&$5.3\times10^{15}$&$1.7\times10^{14}$\\
$L_X$&Extension of the reconnection region (\rm cm)&$1.5\times 10^{14}$&$10^{15}$&$4.4\times10^{15}$&$8.8\times10^{12}$\\
$L$&Extension of the corona (\rm cm)&$3\times 10^{14}$&$2\times10^{15}$&$4.4\times10^{15}$&$8.8\times10^{12}$\\
$V$&Volume of emission region ($\rm cm^3$)&$7.8\times10^{43}$&$2.3\times10^{46}$&$10^{48}$&$1.36\times10^{41}$\\
$d$&Distance of the source(\rm Mpc)&$3.8$&$75$&$16.7$&$78$\\
$m$&Mass of BH ($M_{\odot}$)&$5\times10^{7}$&$3.4\times10^{8}$&$3\times10^{9}$&$10^8$\\
$p$&Injection spectral index&2.4&2.15&2.4&1.7\\
$\gamma_{min}$&Particle minimum Lorentz factor&6&2&4&2\\
$N_H$*&Dust/neutral gas column  density ($\rm cm ^{-2}$)&$10^{23}$&$4\times 10^{20}$&$2\times 10^{20}$&$1.2\times 10^{21}$\\
\hline
\multicolumn{8}{l}{* The observed  values for $N_H$ of Cen A, Per A, M87 and IC310 are taken from}\\
 \multicolumn{8}{l}{\cite{morganti08,canning10,lieu96} and \cite{kalberla10}, respectively.}
\end{tabular*}
\end{minipage}
\end{table*} 

\subsubsection{Photon-neutral ($\gamma N$) interactions}
\label{photoionization}

The low energy  photons produced in the nuclear emission region will  propagate in the surrounding interstellar medium of the host galaxy filled  mainly by hydrogen and helium gas. 
Photons with energies larger than the hydrogen Lyman threshold (13.6 eV) will be able to photo-ionize the neutral gas.

The optical depth resulting from these interactions is approximately given by

\begin{equation} \label{tau-dust}
\tau_{\gamma H}(E_\gamma)=N_H \sigma_{\gamma N}(E_\gamma)
\end{equation}
where  $N_H$ is the neutral hydrogen column density, and  $\sigma_{\gamma N}$ is the absorption cross section. As in \citealt{reynosoetal2011}, we   take this  from \cite{ryter96} for $E_{\gamma}<$ 1 keV considering that atomic hydrogen and galactic dust are the dominant components  of the environment. The values of $N_H$ for each source investigated here are taken from the observations  and are listed in Table~\ref{llagntable}.

This $\gamma N$ absorption has been considered in the reconstruction of  the SEDs of the four LLAGNs  in \S. 3. As we will see there, the photons produced in  the optical-soft X-ray range  are fully absorbed by these interactions.

\subsection{Radiative cooling processes}

In this section we discuss briefly the relevant radiative loss processes for  electrons and protons.

We take into account both leptonic and hadronic radiative loss mechanisms in the emission region. This corresponds to the torus  with volume $V$ that encompasses the cylindrical shell where magnetic reconnection particle acceleration takes place in Figure~\ref{fig1}. Considering that the cylinder extends up to $L$ in both hemispheres, then the small radius of the torus is $r=L/2$ and the large radius is $R_X$, so that the effective emission zone in our model has an approximate volume  $V =  \pi^2 L^2 R_X$.

For leptonic processes, we consider the interactions of relativistic electrons with the surrounding magnetic, charged matter, and  photon fields. 

Accelerated electrons spiralling in the magnetic field emit synchrotron radiation. We calculate the synchrotron  loss rate for the sources considered  here and the  radiated synchrotron spectrum as functions of the scattered photon energy (see Eqs. 10 and 11 in KGV15). 
Electron interactions with the electrostatic field of  nuclei of charge $Ze$ allow for the production of bremsstrahlung radiation. 
Finally, relativistic electron interactions with photons may produce inverse Compton (IC) radiation.
We considered different photon fields in the surrounds of BH, namely, the scattered photons from accretion disk and core black-body radiation (Eq. 36 in KGV15) and the electron synchrotron emission (which allows for SSC; Eq. 14 in KGV15) and found  that the latter, i.e., the SSC mechanism is  dominant  in the inner coronal/accretion disk region  we are interested in this work.

Accelerated protons produce  hadronic emission  from interactions with the  magnetic field (synchrotron), and also through the decay of  neutral pions. These are produced either  by inelastic collisions with nuclei of the corona ($pp$ interactions; Eqs. 19 \& 21 of KGV15), or via interactions with photons, in photo-meson production ($p\gamma$ mechanism; Eqs. 40 \& 43 in KGV15). The latter mechanism takes place for photon energies greater than $E_{th}\approx$ 145 MeV in the reference frame of the relativistic proton. A single pion can be produced in an interaction near the threshold and then decay giving rise to gamma-rays. 
In our model the dominating  photon field  comes from  the synchrotron radiation.
\footnote{We find that  for  photo-meson production, the  radiation from the accretion disk and from the corona are irrelevant compared to the contribution from the synchrotron emission.} 

The cooling rates due to  these leptonic and hadronic mechanisms  are plotted as  functions of the particle energy for the sources here studied in Figures~\ref{2a}, \ref{4a}, \ref{6a} and \ref{ica}, and  Figures ~\ref{2b}, \ref{4b}, \ref{6b} and \ref{icb}, respectively,  where they are also compared with the acceleration rates due to shock and magnetic reconnection.

The radiative losses considered above are used to calculate  the SEDs of the sources 
in  \S 3.

\section{application to radio galaxies}

We describe here  the results of the application of the  model described above  to  Cen A, Per A, M87 and IC 310 which are classified as radio galaxies. 
As remarked in \S. 1, these radio galaxies have been observed at VHE  by \textit{FERMI-LAT, VERITAS \rm{and} $HESS$} (e. g., \citealt{abdo09,abdo10,abram12,aleksic14}).  

In general lines, in  all cases, the procedure to calculate the SED begins with the determination of the total power released by fast magnetic reconnection  within the acceleration region (Eq.~\ref{2}), that is, in the cylindrical shell of height $L_X$ and thickness $\Delta R_X$  (Eq.~\ref{3}) (see Figure 1). This power is then employed to compute the spectrum of accelerated electrons and protons (Eq.~\ref{10}) that will be injected into the emission volume $V$, i.e., the torus that surrounds the acceleration region (as described in \S. 2.5). The parameters employed for each source are given in Table 1.
We note that our model has actually only 7 free parameters (i.e., $R_x$, $L_X$, $L\leq L$, $p$, $\gamma_{min}$, $\xi$ and $\alpha$). The remaining quantities of Table 1 are obtained directly from these parameters through Eqs. 1 to 5 (i.e, $B$, $W$, $\Delta R_X$, $n_c$ and $T_d$), or from the observations (i.e, $d$ and $m$).
 The maximum energy that each particle spectrum can attain is obtained from the comparison of the acceleration rates, for both electrons and protons, with the relevant radiative loss processes (seen \S. 2). As remarked, the accelerated electrons will loose energy by synchrotron, IC and Bremsstrahlung  mechanisms, with a dominance of the synchrotron process shaping their spectrum. The fluxes of these emission processes are then calculated and also the number density of the synchrotron photons that are partially self-scattered by the electrons (leading to  SSC emission)  and  by protons (in $p\gamma$ interactions).
Likewise,  the energy distribution of the protons is also calculated taking into account the radiative cooling mechanisms due to synchrotron, $pp$ and $p\gamma$ interactions (see \S 2.3) that will shape the very high energy part of the SED.


 \subsection{Application to Cen A}

The Prominent radio galaxy Cen A (or NGC 5128) is the nearest FR I active radio galaxy to Earth (z=0.0018, \citealt{graham78}), at a distance of $\simeq$ 3.8 \rm{Mpc} (\citealt{rejkupa04}), making it uniquely observable among this class of objects and an excellent source for studying the physics of relativistic outflows as well as of the core region. Cen A is one of the best well known extragalactic objects over a wide range of  frequencies and the photon emission from the nuclear region of the galaxy has been detected from the radio  to the $\gamma$-rays band. Cen A has been proposed as a possible source of UHE cosmic rays (with energies $\leq 6\times  10^{19} \rm{eV}$; \citealt{abraham07}) by the Pierre Auger collaboration. 
The SMBH mass inferred from kinematics of stars, as well as $H_2$ and ionized gas is estimated to be in the range of $\sim10^7-10^8\ M_{\odot}$ (\citealt{marconi06,neumayer07}), and here we adopted the value $5\times 10^7 M_{\odot}$. 
The  viewing angle of the jet ($\theta$) is still debatable, for instance at parsec scales it is  $\theta \sim 50^{\circ}-80^{\circ}$ (\citealt{tingay98}), whereas at the 100 pc scale $\theta \sim 15^{\circ}$ (\citealt{hard03}). 

In this section we show the results for Cen A  obtained by applying the model described in \S. 2  around the nuclear region, employing   the set of parameters listed in Table 1.

The values for the first five parameters in Table 1 have been calculated from Eqs.~\ref{1}-\ref{4} and \ref{11}. We take for the accretion disk inner radius the value $R_X=6 R_S$,  for the extension $L_X$ of the reconnection region (see Figure~\ref{fig1}), we consider the value $L_X \simeq 10R_S$, and for the extension of the corona $L \simeq 20R_S$.  
As remarked earlier, the volume $V$ of the emission region in Table 1 was calculated by considering the torus that encompasses the reconnection region in Figure~\ref{fig1}. The magnetic reconnection power $W$ is evaluated from Eq.~\ref{2}.

\begin{figure}
    \centering
    \subfigure[]
    {
        \includegraphics[width=3in]{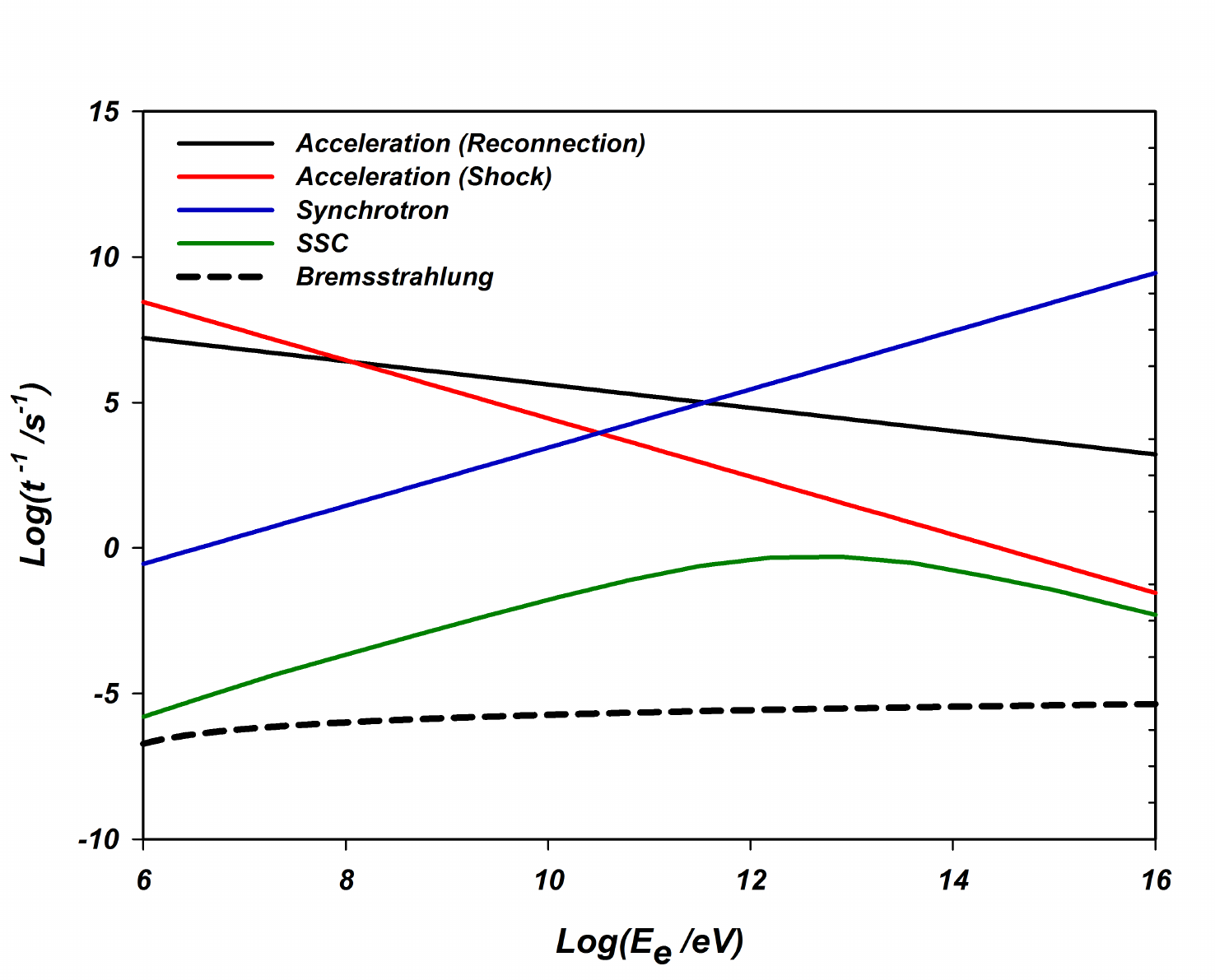}
        \label{2a}
    }
    \\
    \subfigure[]
    {
        \includegraphics[width=3in]{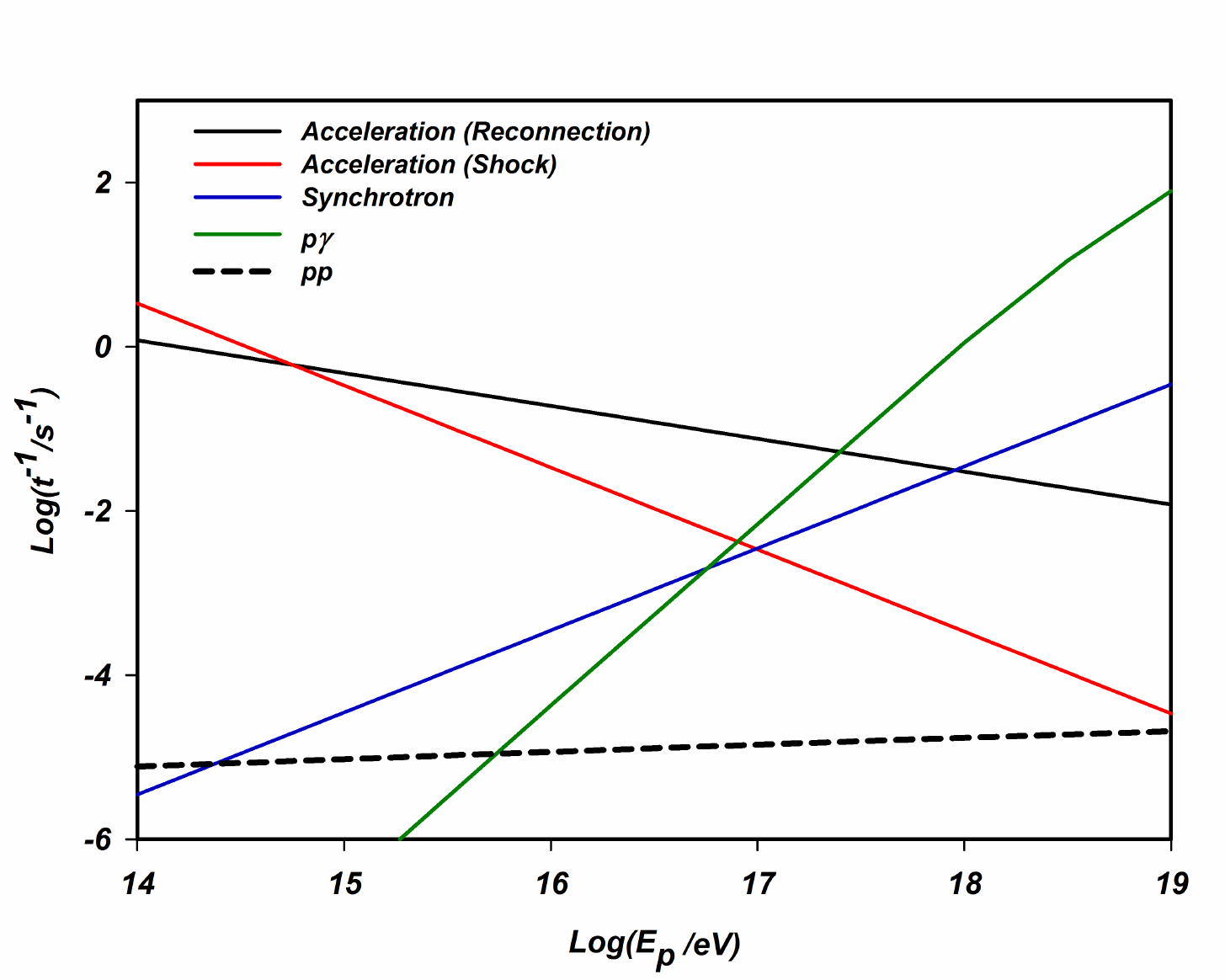}
        \label{2b}
    }
       \caption{Acceleration and cooling rates for electrons (a) and protons (b) in the nuclear region of Cen A.}
    \label{fig2}
\end{figure}

\begin{figure}
  \centering
  \includegraphics[width=0.50\textwidth]{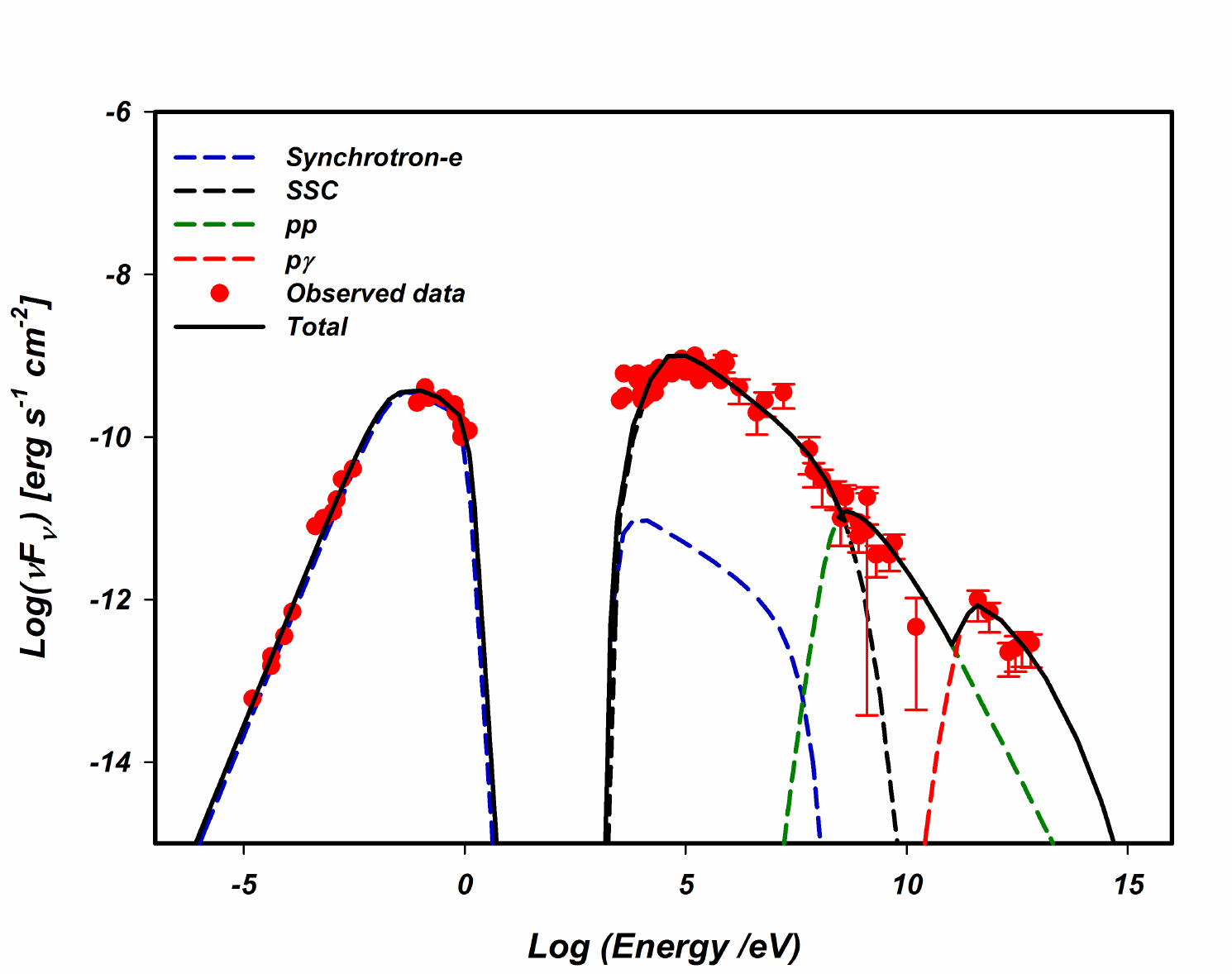}
  \caption{Calculated spectral energy distribution (SED) for Cen A using the magnetic reconnection acceleration model in a lepton-hadronic scenario compared with observations. 
  The data depicted in the radio to optical energy range  ($10^{-5}\rm{eV}- 1 eV$) is from SCUBA at 800 $\mu m$ (\citealt{hawarden93}), ISO \& SCUBA at 450 $\mu m$ and 850 $\mu m$ (\citealt{mirabel99}); the data in the  hard x-rays range is  from \textit{Swift-BAT} (\citealt{ajello09}) and \textit{Suzaku} (\citealt{marko07}). We also include data from \textit{OSSE} (\citealt{kinzer95}) and \textit{COMPTEL} (\citealt{steinle98}) in the range of $5\times 10^5-10^7\rm{eV}$. The data observed in the energies $10^8-10^{10}\rm{eV}$ are taken by \textit{EGRET} (\citealt{sreekumar99,hartman99}) and in the energies $10^8-10^{10}\rm{eV}$ by \textit{Fermi-LAT} (\citealt{abdo09a,abdo10}). The \rm{TeV}\  data  are taken by \textit{HESS}  (\citealt{ahar09}).}
 \label{3a}
\end{figure}
 
Figure~\ref{fig2} shows the radiative cooling rates for the different energy loss processes for electrons and protons as described in Section 2.4. These are compared with the acceleration rates due to first-order Fermi acceleration both within the magnetic reconnection site (Eqs.~\ref{5} \&~\ref{6}) and  behind a shock (Eq.~\ref{7}). We notice that at high energies for both protons and electrons the acceleration is dominated by the first-order Fermi magnetic reconnection process in the core region. Besides, the main radiative cooling process for  the electrons is synchrotron radiation (Figure~\ref{2a}), while for protons the photo-meson production ($p\gamma$ interactions) governs the loss mechanisms (Figure~\ref{2b}). For the $p\gamma$ interactions, we have found  that the proper target radiation field is that of the photons from the electron synchrotron emission.

The intercept between the magnetic reconnection acceleration rate and the synchrotron rate in Figure~\ref{2a} gives the maximum energy that the electrons can attain in this acceleration process, which is  $\sim 3\times 10^{11} \rm{eV}$. 
Protons on the other hand, do not cool as efficiently as the electrons and can attain energies as high as $\sim2.5\times 10^{17}\rm{eV}$.

We have constructed the SED of Cen A using a lepton-hadronic model where particles are accelerated close to the central BH by magnetic reconnection and interact with the surrounding fields   radiating  in a spherical region of radius $L$ (see section 2.4). The SED is depicted in (Figure~\ref{3a}) where it is compared with the observations. 

We note that the data presented in this figure in the low and intermediate energy ranges come  mostly from archival data  and represent typical average emission and activity. 
The emission in the high energy range obtained by \textit{Fermi-LAT} and \textit{HESS} at the same epoch  correspond to non-variable and moderate activity too.

As described in section 2.3, we have adopted a particle energy injection power law function:
\begin{equation} \label{13}
Q(E)\propto E^{-p}.
\end{equation}
 In Figure~\ref{3a}, we considered injected primary particles with $p=2.4$ which is consistent with theoretical predictions of particle acceleration within magnetic reconnection sites  (see \S. 2.3).
 Our calculations show that synchrotron radiation explains the observed  emission in the radio to visible band, while SSC is the dominant mechanism to produce the observed hard $X-$rays and low energy  $\gamma$-rays  as a result of  interactions between energetic electrons with scattered synchrotron photons. Also in the Figure,  neutral pion ($\pi^0$) decays can explain the observed $\gamma$-rays at TeV energies, via $pp$ and $p\gamma$ interactions which are  the two main processes producing  $\pi^0$.

We note that in order to fit the observed data in the radio to optical range, we had to assume a minimum energy for the injected electrons  in the acceleration zone (Eq. 11 in KGV15),  $E_{min}=(\gamma_{min}-1) m_e c^2$, with $\gamma_{min} =6$. Though this injected value has no influence on the VHE tail of the SED, it is determinant in the match of the low energy branch. We have found that values of $\gamma_{min} <6$ do not lead to the synchrotron match in the low energy range   (see also \S 4).

As remarked in \S. 2.4, the $\gamma$-ray absorption due to pair production occurs according to  Figures~\ref{8a} and \ref{8b} very near  the accretion disk at heights smaller than $\sim 0.001R_S$, thus much smaller than the emission region that extends up to $\sim20R_S$ in our model, so that $exp(-\tau)\simeq1$ and the absorption effect is not effective at the heights of  interest. 
On the other hand, due to the high dust and neutral gas column  density in Cen A ($N_H=10^{23} \rm {cm^{-2}}$, see e.g., \citealt{morganti08}) we find that the  optical to soft X-ray emission is fully absorbed via $\gamma N$ absorption (see figure~\ref{3a}, Eq.~\ref{tau-dust}).

 \subsection{Application to Per A }
Perseus A (also known as NGC 1275 and 3C 84), is a nearby active galaxy located at the centre of the Perseus cluster and hosts a central SMBH mass of $\sim 3.4\times10^8\rm{M_\odot}$ (\citealt{wilman94}). In fact, Per A is one of the closest $\gamma$-ray emitting AGNs. Its distance to the Earth is 75 \rm{Mpc} (\citealt{brown11}) and is also of great interest, specially due to its proximity, also providing  an excellent opportunity to study the physics of relativistic outflows. Per A also seems to exhibit  jet precession with an orientation angle $\approx 30^{\circ}-55^{\circ}$ (\citealt{walker94}; see also \citealt{falceta10}), which may be an indication that Per A is the result of a merger between two galaxies (\citealt{liuchen07}). It is a very bright radio galaxy showing an extended jet with FR I morphology (e.g., \citealt{verm94,butt10}) with asymmetric jets at both \rm{kpc} (\citealt{pedlar90}) and \rm{pc} scales (\citealt{asada06}). 

The  parameters of our model for producing the SED of  Per A are tabulated in Table 1.  
The first five parameters are calculated from Eqs.~\ref{1}-~\ref{4} and \ref{11}. 
As for Cen A, we  have also used for the accretion disk inner radius the value $R_X = 6 R_S$ and for the extension $L_X$ of the reconnection region the values $L_X=10 R_S$ and $L\simeq20 R_S$.

\begin{figure}
    \centering
    \subfigure[]
    {
        \includegraphics[width=3in]{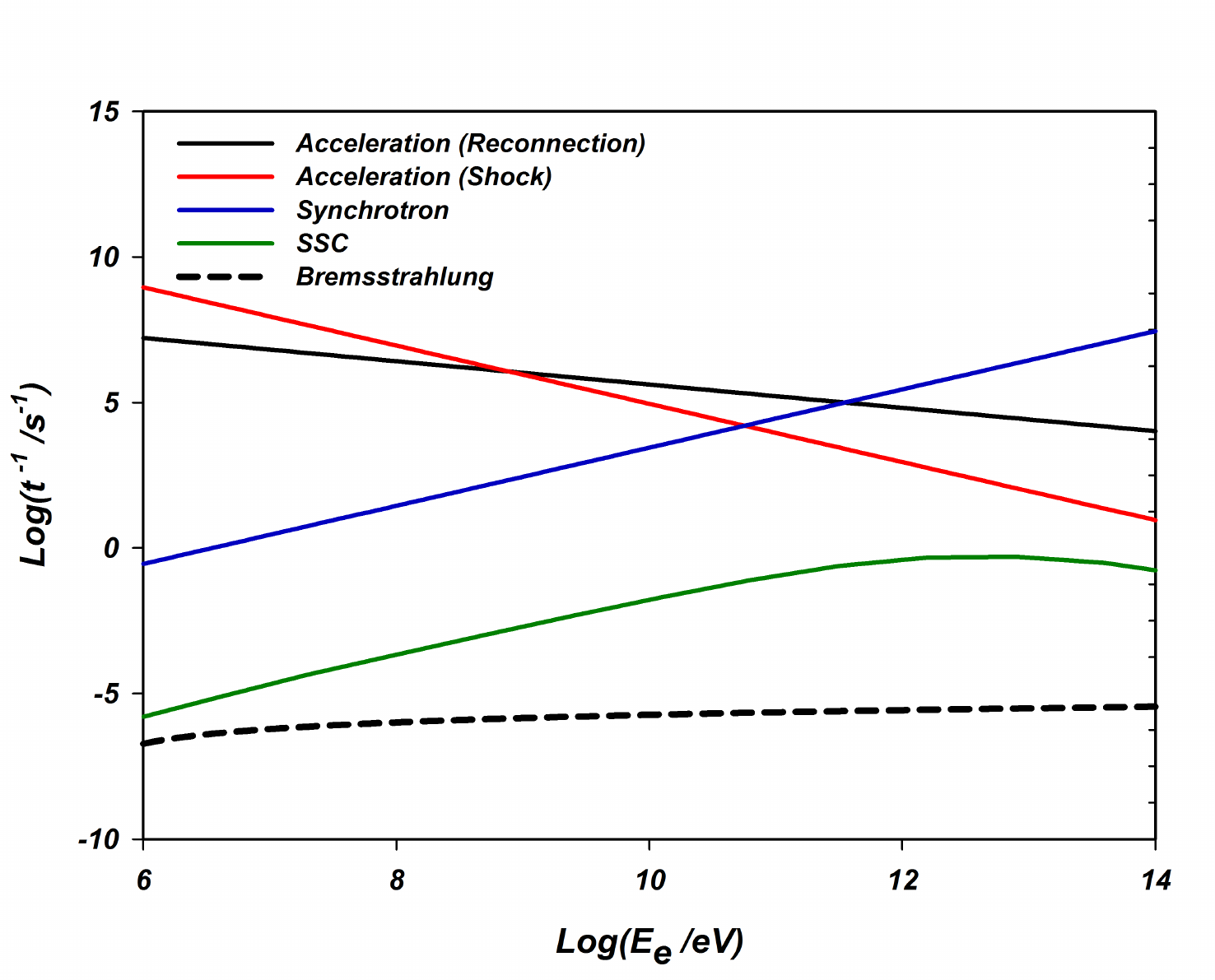}
        \label{4a}
    }
    \\
    \subfigure[]
    {
        \includegraphics[width=3in]{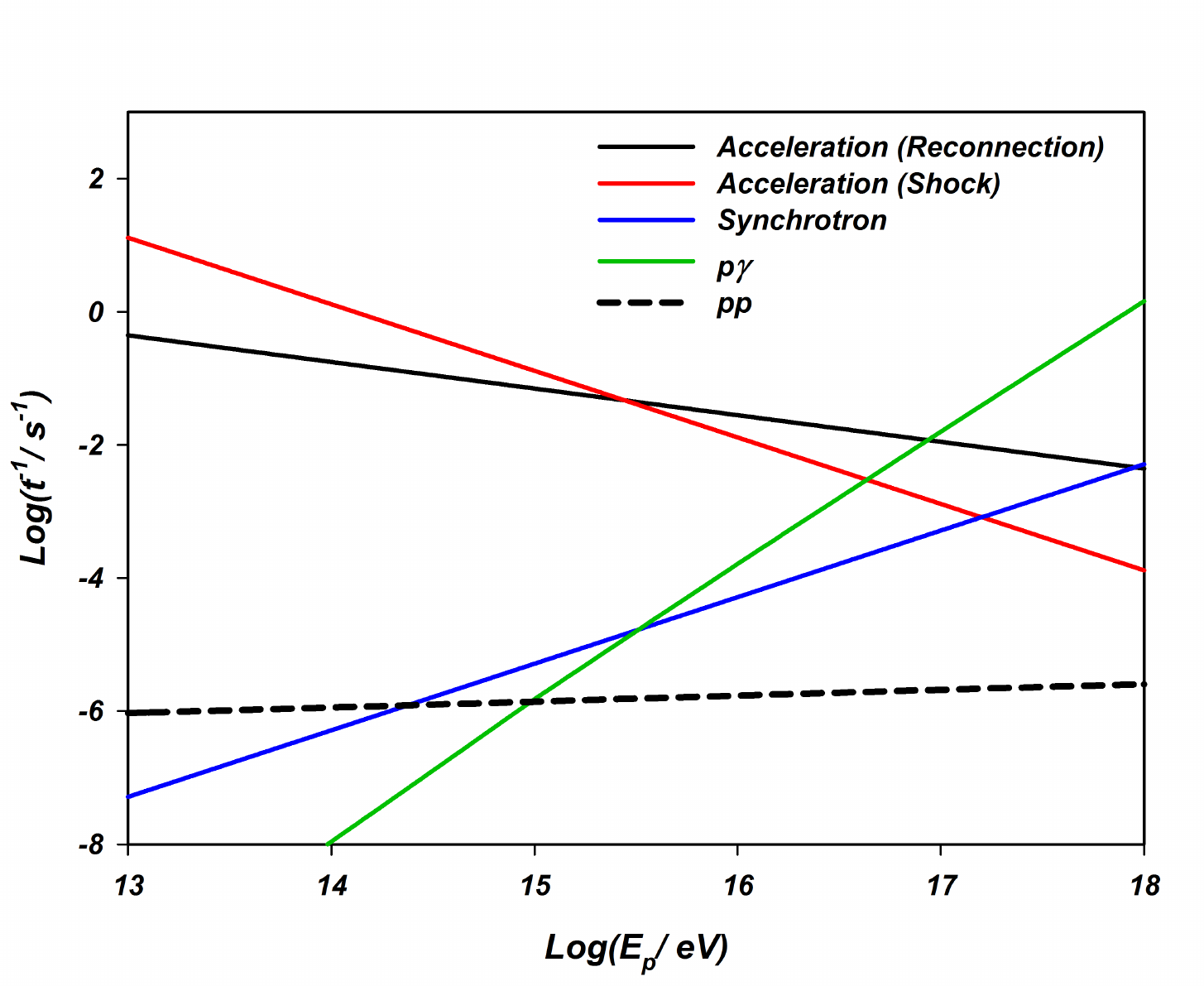}
        \label{4b}
    }
   
    \caption{Acceleration and cooling rates for electrons (a) and for protons (b) in the nuclear region of Per A (NGC 1275).}
    \label{fig4}
\end{figure}

The radiative loss and acceleration rates for electrons and protons are compared in Figure~\ref{fig4}.  As in Cen A,  magnetic reconnection is the dominant acceleration mechanism over shock acceleration at the high energy branch  for both electrons and protons and determines the maximum energy that the particles can achieve before losing part of it radiatively. Electrons  may be accelerated   up to $3\times 10^{11}\rm{eV}$ and the main process to cool them is synchrotron. While the maximum energy the  protons can achieve is $10^{17}\rm{eV}$ and \textit{photo-meson} production ($p\gamma$) is the dominant mechanism to cool them. Similarly to Cen A, the dominant photon field  interacting  with the accelerated protons is the synchrotron radiation.  
  
 \begin{figure}
  \centering
  \includegraphics[width=0.450\textwidth]{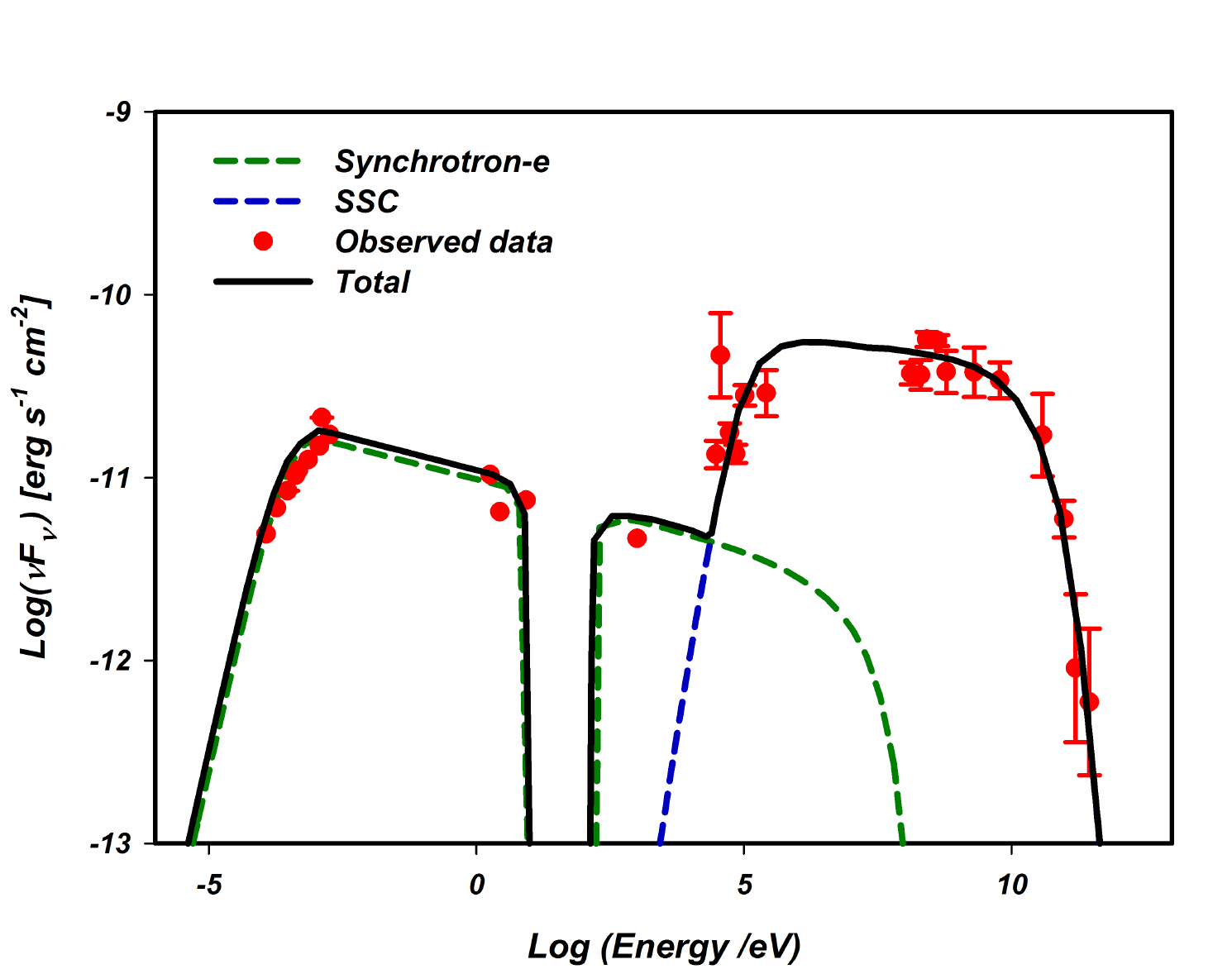}
  \caption{A leptonic model to reproduce  the SED of Per A (NGC 1275) using the magnetic reconnection acceleration model. Data include \textit{MOJAVE} (\citealt{lister09}), \textit{Planck} (\citealt{ade11}), HST (\citealt{chiaberge99}), and HST FOS (\citealt{johnstonefabian95}) for the radio to optical spectrum;  data depicted in X-rays is from the XMM (\citealt{torresi12}), \textit{Swift}-BAT (\citealt{ajello09}), and BATSE (\citealt{harmon04}); and data depicted in the gamma-ray band is from \textit{Fermi-LAT} (\citealt{abdo09c,ackermann12}) in the 100 \rm{MeV}-100\rm{GeV} energy range, and from \textit{MAGIC} (\citealt{aleksic10a,aleksic10b,aleksic12a,aleksic12b}) in the VHE tail. We note that the error bars for the BATSE data (in the $10^5$ eV range) were evaluated using \cite{harmon04,soldi14} and \cite{wilson12}.}
 \label{5a}
 \end{figure}

  We have constructed the SED for this source employing  a leptonic scenario (Figure~\ref{5a}).  
 In this case,  the primary particles were injected with a power law spectral index $p=2.15$ (Eq.~\ref{8}). The radio spectrum is matched by electron synchrotron emission, with particles injected into the acceleration zone with  rest mass energy (i.e., with $\gamma_{min} =2$). The observed  X-ray and $\gamma$-ray emission is nearly reproduced by SSC occurring in the nuclear region in a spherical region of radius $L \sim 20 R_S$, as described in \S. 2.4. 
However, it should be noticed that there is a high intensity BATSE data point in the $\sim 10^5$ eV  that is not  matched by the model. Given the fact that this source is highly variable,  this feature probably corresponds to a more active state superposed to the less active one (represented by the less intense data points in the same energy range). 
 
 The  observations also indicate that there is a high energy cut-off around $\sim 3\times 10^{11}$eV in this source. In our scenario this is due  to  leptonic emission produced by interactions of high energy electrons with the radiation field produced by themselves and this cut-off is compatible with the maximum energy calculated from the comparison of the reconnection acceleration rate with the synchrotron loss rate  in Fig. \ref{4a}.

As stressed in \S. 2.5, the optical depth for the produced $\gamma$-rays was also calculated in this case and is shown if  Figures~\ref{8c} and \ref{8d}).  We note that the 100 GeV $\gamma$-rays may be fully absorbed due to pair production but only very near to disk ($z<0.001R_s$). However, these vertical distances from the disk, comparing to the length scale of the emission region is very small and reasonably, we can ignore the absorption effect at the heights larger than $\sim R_s$ which is compatible with extension of emission region in our model.

Similarly to Cen A,  the neutral gas and  dust of the interstellar medium of the host galaxy in this source also causes the extinction of  the emission in the range of $10-10^2 \rm{eV}$  (Fig.~\ref{5a}; see also \S.~\ref{photoionization}).

 \subsection{Application to M87}
 \begin{figure}
    \centering
    \subfigure[]
    {
        \includegraphics[width=3in]{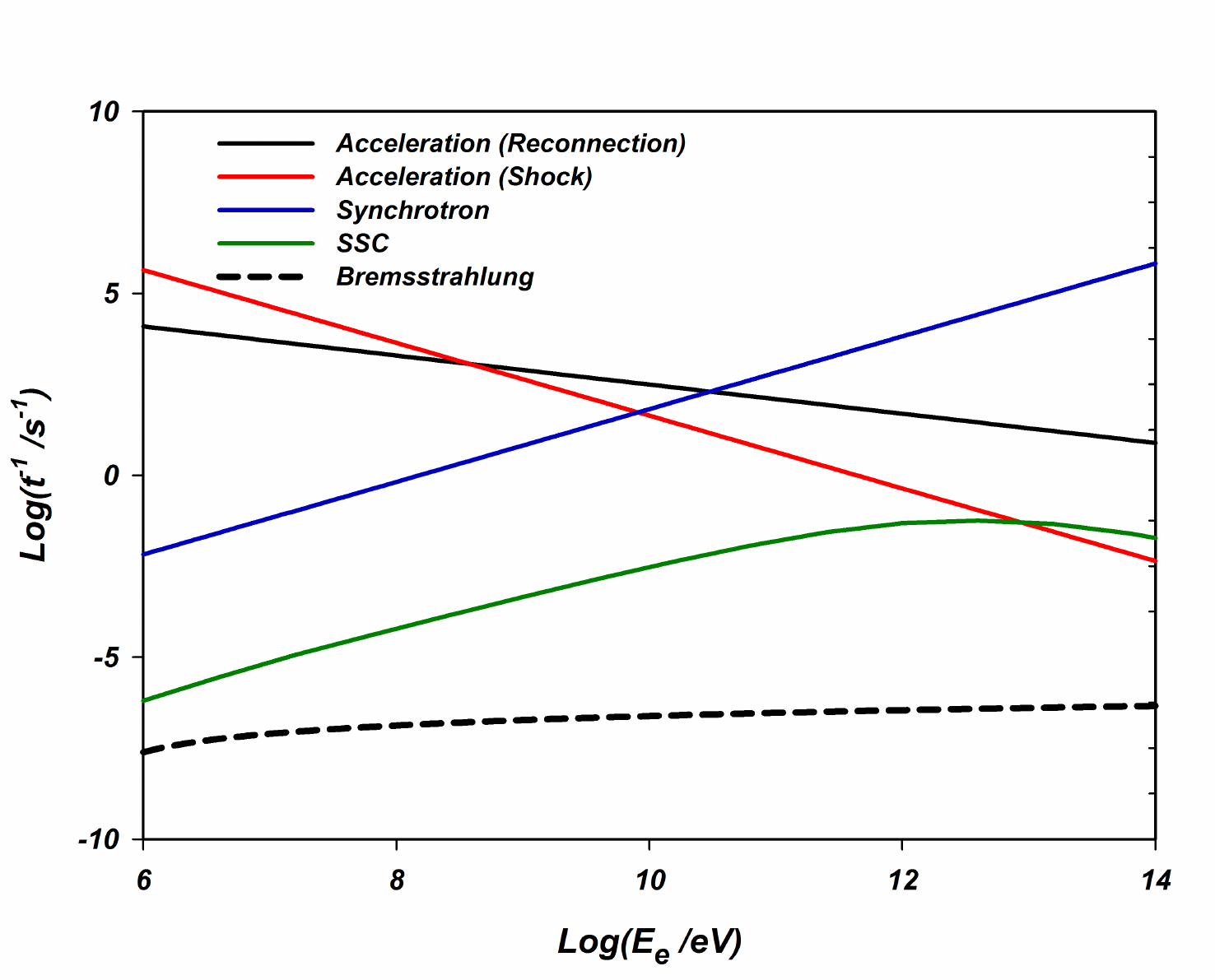}
        \label{6a}
    }
    \\
    \subfigure[]
    {
        \includegraphics[width=3in]{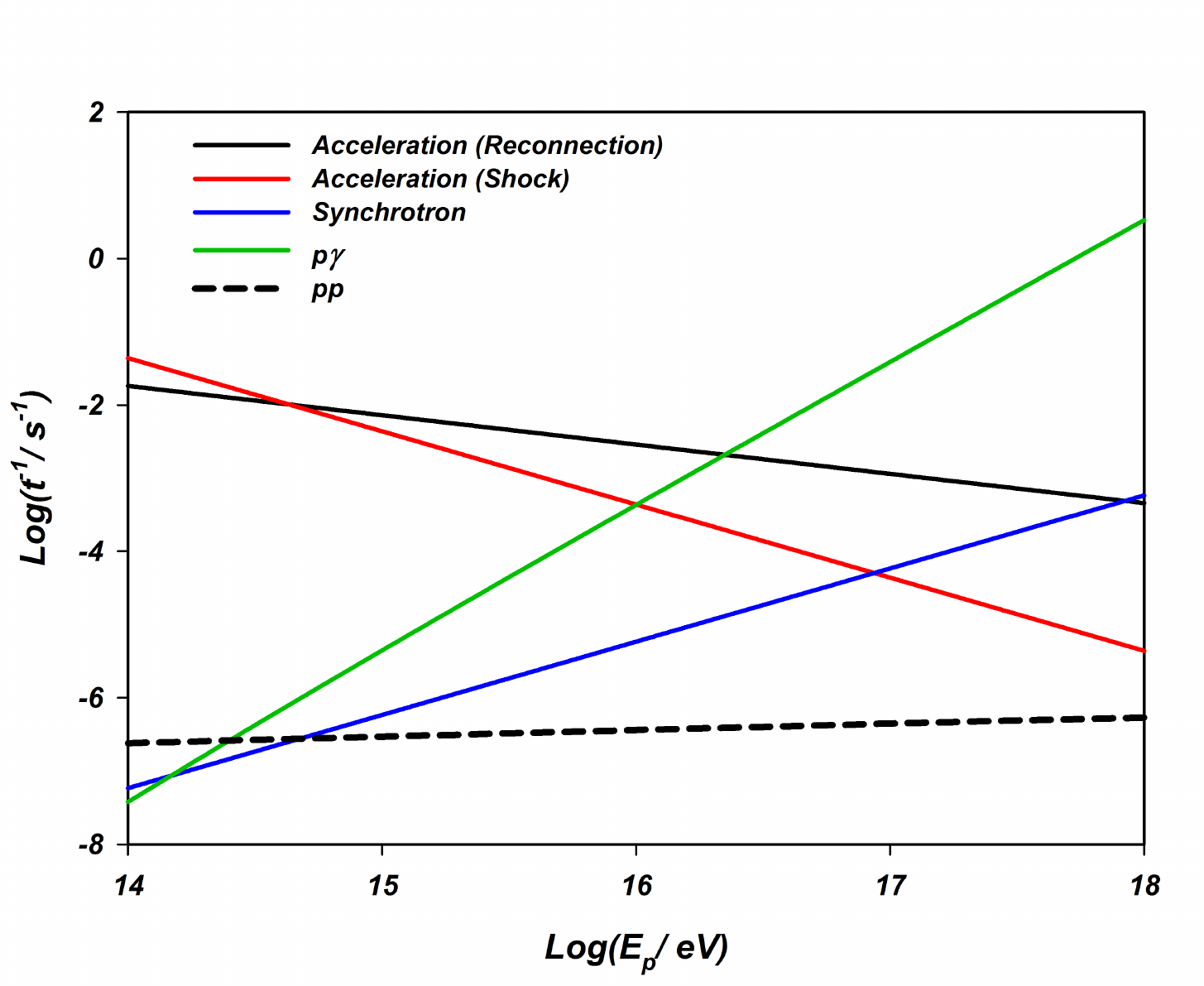}
        \label{6b}
    }
   
    \caption{Acceleration and cooling rates for electrons (a) and for protons (b) in the nuclear region of M87.}
    \label{fig6}
\end{figure} 
 The FR I giant radiogalaxy M87 is another  well-known nearby AGN located at 16.7 \rm{Mpc} within the Virgo cluster which harbours a SMBH with a mass of $M_{BH}\sim 6\times10^9\ M_{\odot}$ (e. g., \citealt{geb09}) which, along with Cen A and Per A, has been known as a peculiar extragalactic  laboratory to study  high energy astrophysics and investigate the nonthermal mechanisms of VHE emission in AGNs. 
 The observations indicate that its jet is oriented within $20^{\circ}$ of the line of sight (\citealt{biretta99}), so that as in the other cases, no significant Doppler boosting is  expected for the $\gamma$-ray flux.  
  
The \rm{TeV} $\gamma$-ray signal from M87 was first reported by \textit{HEGRA} (\citealt{ahar03}) and then confirmed by \textit{HESS} (\citealt{ahar06}). The latter  also revealed that this emission is strongly variable with time scales of 1-2 days and thus produced in a very compact region, as pointed out before.

 Table 1 shows the parameters that we used to calculate the acceleration and cooling rates and also to reconstruct the SED of this source.

In Figures~\ref{fig6}, we compare the rates of the radiative cooling processes with the rates of the acceleration mechanisms for electrons and protons. As in the other two sources, we find that the  dominant energy loss  mechanisms  are the synchrotron and the $p\gamma$ interactions for electrons and protons, respectively, and  the acceleration is dominated by the magnetic reconnection process which  defines the energy cut off for both electrons and protons. Figure~\ref{6a} indicates that this maximum energy  is $\sim4\times10^{10}\rm{eV}$ for electrons and $\sim5\times10^{16}\rm{eV}$ for protons.

 \begin{figure}
  \centering
  \includegraphics[width=0.50\textwidth]{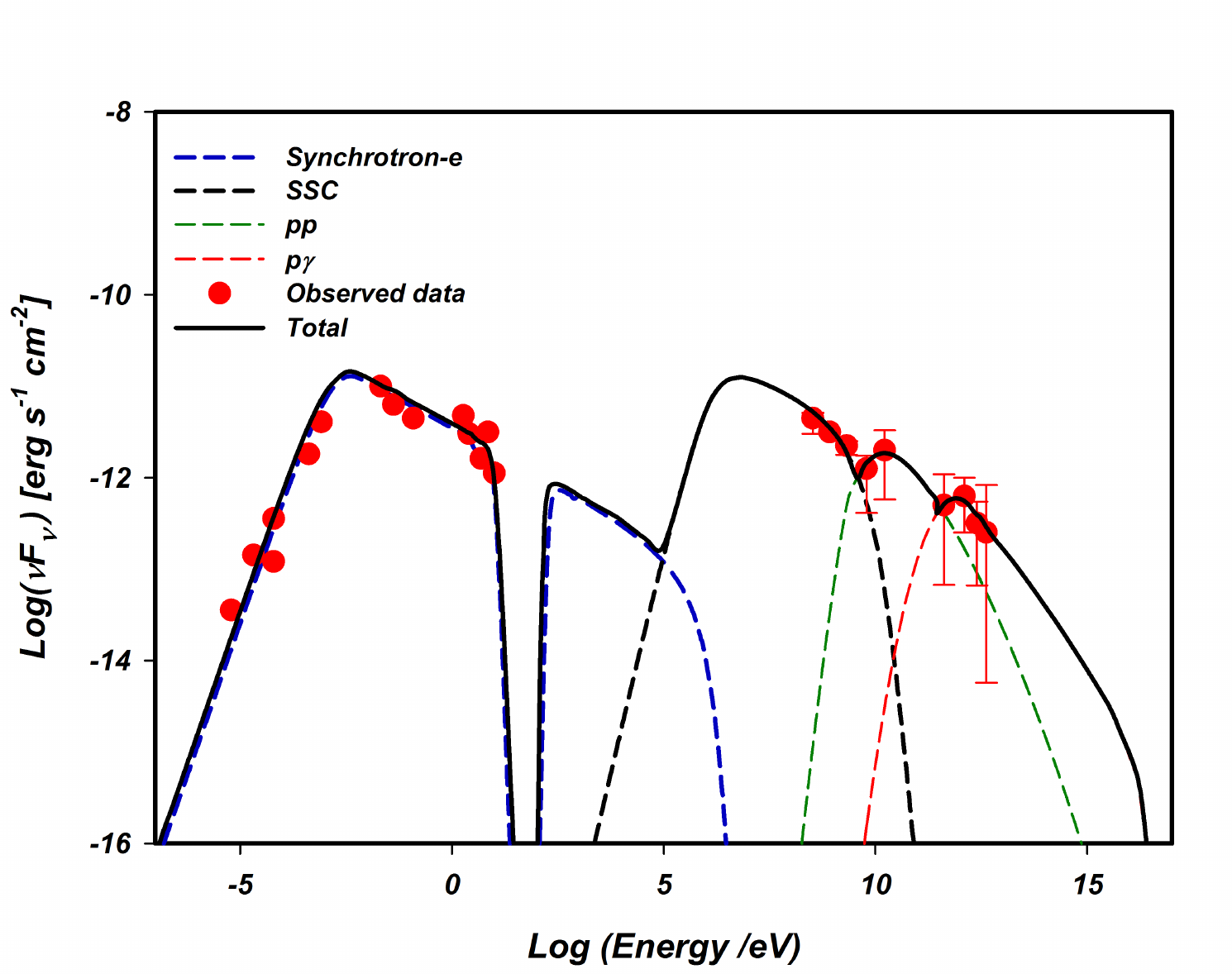}
  \caption{A lepton-hadronic model of the SED of M 87  compared with observations. The core radio data are obtained from MOJAVE VLBA (\citealt{kellermann04}) at 15 GHz, from (\citealt{biretta91})  at 1.5, 5 and 15 GHz, from IRAM (\citealt{despringre96}) at 89 GHz, from SMA at 230 GHz (\citealt{tan08}), from Spitzer at 21 and 7.2 GHz (\citealt{shi07}) and from Gemini (\citealt{perlman01}) at 3.2 GHz. Optical-UV emission from HST  (\citealt{sparks96}).  \rm{MeV/GeV} $\gamma$-ray  data are  from \textit{Fermi-LAT} (\citealt{abdo09}), and the  low-state TeV spectrum (\citealt{ahar06}) from \textit{HESS}.  
   }
 \label{7a}
 \end{figure}

 Figure~\ref{7a} shows the calculated  SED for M87  compared to the observations. It is also reproduced by a lepton-hadronic model in the core region as described in \S. 2, where we assumed an injected particle energy distribution   $\propto E^{-p}$ with a power index  $p=2.4$. 
 
With  an electron minimum energy  $E_{min}=4 m_ec^2$, we can fit the observed  core radio to visible spectrum by synchrotron emission.

As in Cen A, the low and intermediate energy data in  Figure~\ref{7a} come from archive and represent typical average emission. The data obtained by \textit{Fermi-LAT} ($10^8-10^{11}$eV) and   by \textit{HESS} (the  TeV tail) correspond both to more quiescent states and have been taken in different epochs. They are  reproduced in our model by different mechanisms. While  \textit{Fermi-LAT} data are fitted by SSC and $pp$ collisions,  \textit{HESS} data are fitted by the decay of neutral pions from $p\gamma$ interactions with  photons coming from the synchrotron radiation. We note that an observed flare state by \textit{HESS} (Aharonian et al.
2006; not shown in Fig.~\ref{7a}) can be also reproduced  by our model assuming a flatter injection particle spectrum with a spectral index = 2.1).

Figures~\ref{8e} and \ref{8f} show the absorbed $\gamma-$ray flux for M87. As in the other cases, this absorption is significant only  at heights smaller than $R_S$ and therefore, its effect can be neglected at the much larger emission scales considered here. 

The absorption of low energy photons by interstellar neutral gas and dust in this source is also important (Fig.~\ref{7a}).
 
We note that our  model and the chosen parametrization is also consistent with  the observed TeV rapid variability of  M87 which is $\sim 1-2$ days (\citealt{abram12})  implying an extremely compact emission region (corresponding to scales of only a few $R_S$).
As remarked in \S 2.4, the emission region in our model corresponds to the torus region  that encompasses the cylindrical shell where magnetic reconnection particle acceleration takes place in Figure~\ref{fig1}, i.e.,  the effective emission zone in our model has a thickness $\simeq L$.    For this source $L \simeq 5 R_S$ (see Table 1), which is of the order of  the inferred  scale from the observed variability.

\subsection{Application to IC 310 }

\begin{figure}
    \centering
    \subfigure[]
    {
        \includegraphics[width=3in]{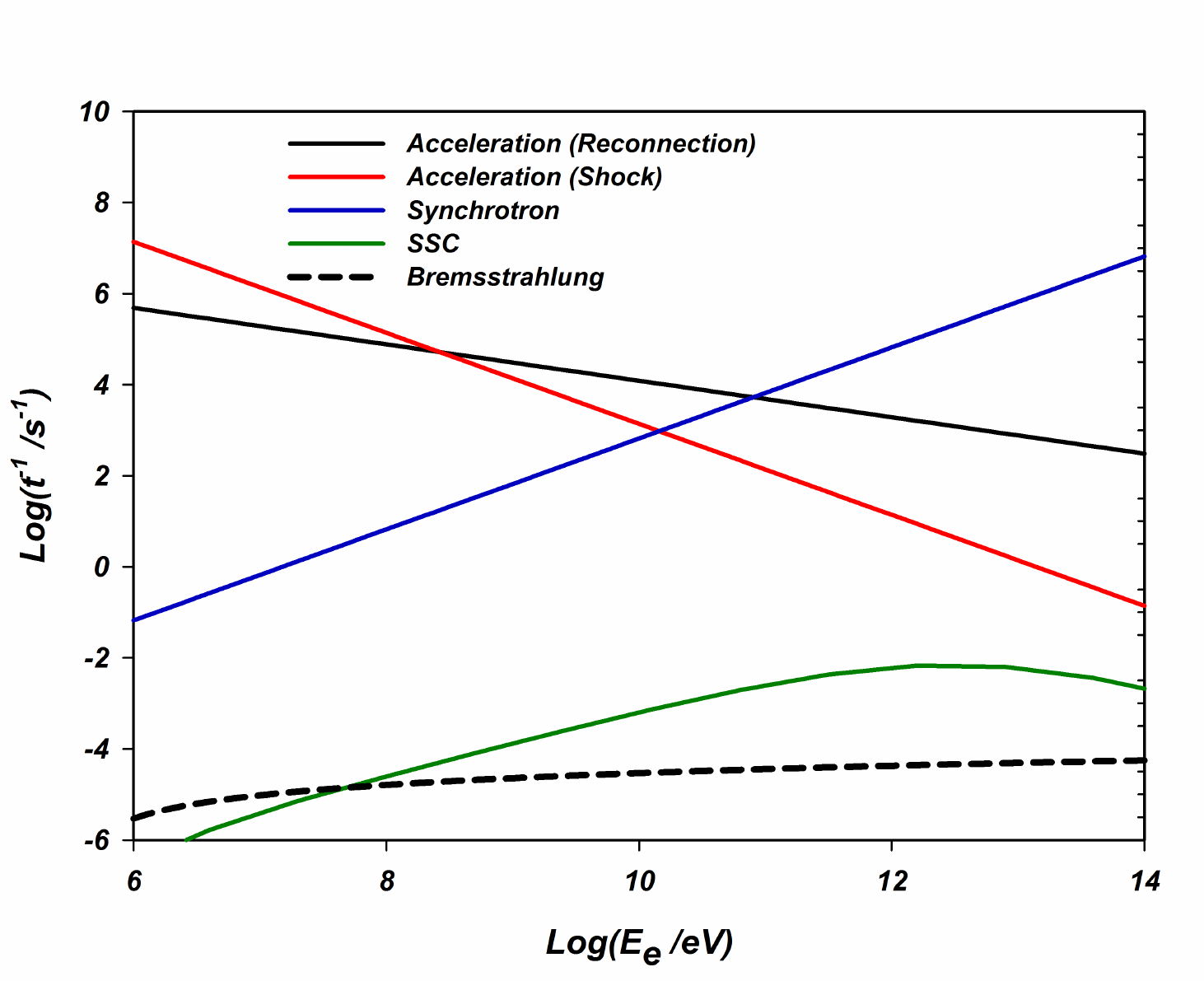}
        \label{ica}
    }
    \\
    \subfigure[]
    {
        \includegraphics[width=3in]{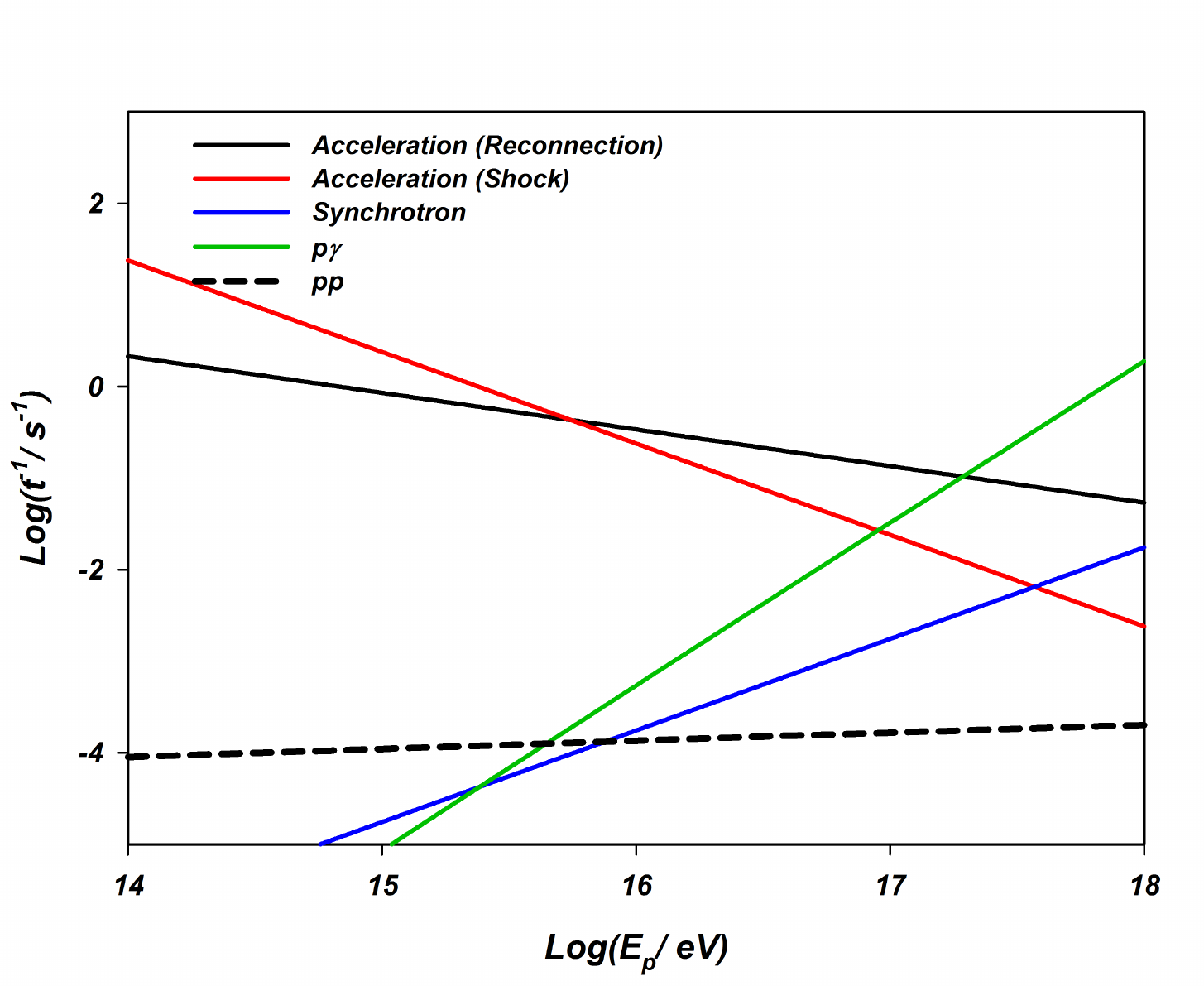}
        \label{icb}
    }
   
    \caption{Acceleration and cooling rates for electrons (a) and for protons (b) in the nuclear region of IC 310.}
    \label{ic}
\end{figure} 

The  peculiar galaxy IC 310 (also named  B0313+411 and J0316+4119 in observational reports) is one of the brightest objects which, as Per A, is also located in the Perseus galaxy cluster at a distance of 78 Mpc from Earth (\citealt{aleksic14b}) and harbours  a supermassive BH with a mass of $\sim 10^8 M_{\odot}$ (\citealt{aleksic14b}). The redshift of this source is z=0.0189 (\citealt{bernardi02}) which has made it the fourth nearest AGN at VHE gamma-rays (\citealt{kadler12}), after Cen A with z=0.00183, M 87 with z=0.004 and Per A with z=0.017559.

IC 310 has been observed  at energies E$>$100 GeV by MAGIC (\citealt{mariotti10}) and  Fermi-LAT collaboration also reported the detection of photons above 30 GeV (\citealt{neronov10}). However the origin of the gamma-ray emission is not clear yet and both  the jet and the core have been considered as  possible emission regions.  

Recently, MAGIC collaboration has reported fast time variability for IC 310 on the VHE $\gamma$-ray with time scales $\sim$4.8 min (\citealt{aleksic14c,aleksic14b}) which constrains the size of the emission zone to  $20\%$ of its $R_S$. 

The parameters we used to calculate the acceleration and cooling time scales and also to reconstruct the SED of this source are shown in Table 1.

The comparison between the acceleration  and cooling rates are depicted in Figures \ref{ica} and  \ref{icb} for electrons and protons, respectively. As in the other cases, we see that  the calculated maximum energy for both electrons and protons reaches larger values for magnetic reconnection than for shock acceleration, so that magnetic reconnection should be the dominating mechanism to accelerate particles in the nuclear region of this source as well. The diagrams indicate that electrons can accelerate up to $8\times 10^{10}$ eV, while the protons up to $2\times 10^{17}$ eV. Also in this source  synchrotron emission is the dominant loss mechanism for electrons and $p\gamma$ radiation is the dominant one for protons for energies larger than $\sim 10^{15}$ eV.

\begin{figure}
  \centering
  \includegraphics[width=0.50\textwidth]{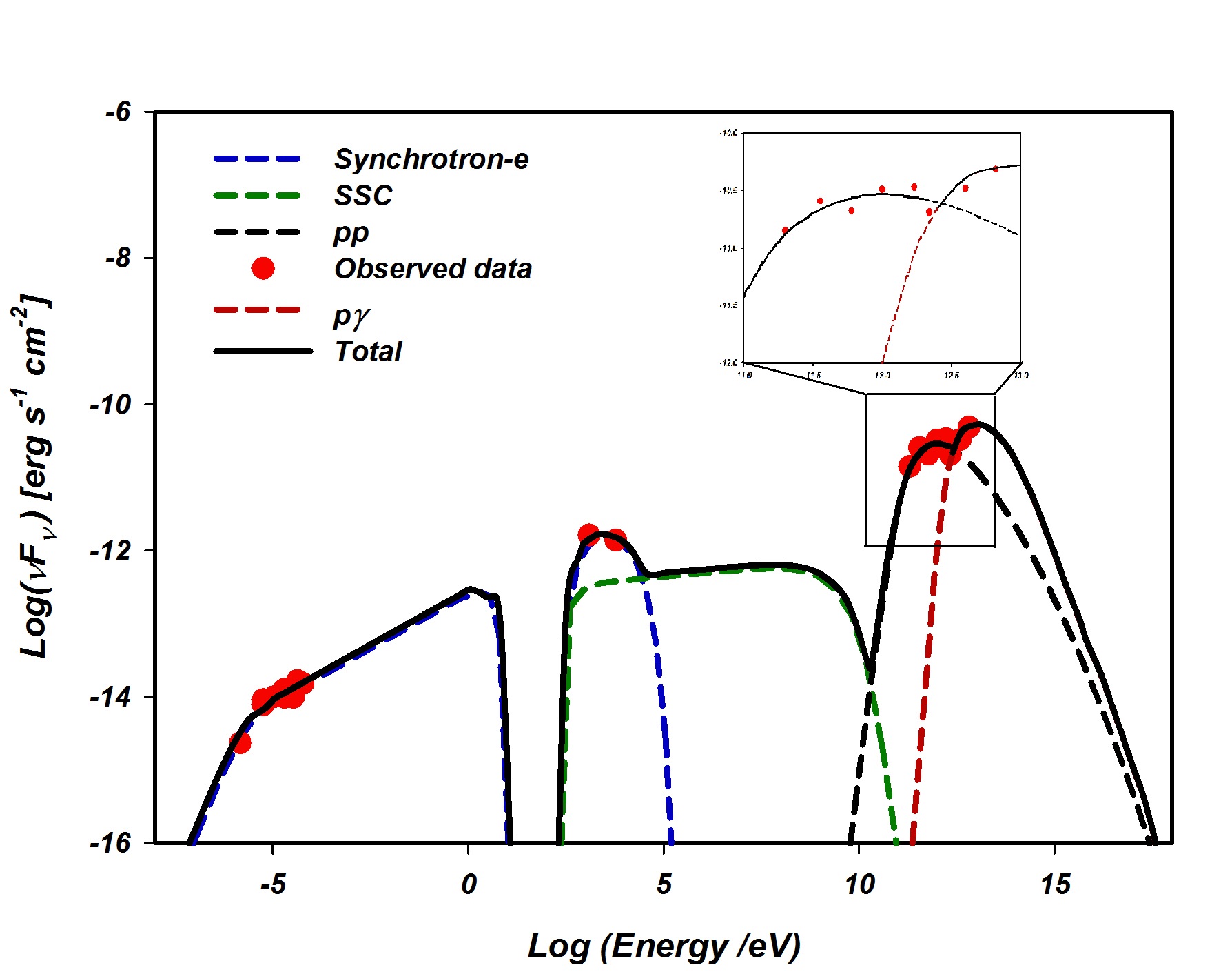}
  \caption{A lepton-hadronic model of the SED of IC 310 compared with observations. The core radio data are obtained from \citealt{kadler12,dunn10,becker91,white92,condon02,douglas96}. The X-ray data are from XMM-Newton (\citealt{sato05}) and the VHE $\gamma$-ray  data are  from \textit{MAGIC} (\citealt{aleksic14b}).  In the upper right side of the diagram it is depicted the detail of the modeling of the VHE branch.
   }
 \label{icsed}
 \end{figure}

Figure \ref{icsed} shows the calculated SED for  IC 310. As in the other sources, the observed radio emission can be explained by synchrotron and the TeV $\gamma$-rays by the  $pp$ and $p\gamma$ processes due to  particles injected with a power law spectral index p=1.7.
 The core opacity to this emission   has been also calculated for IC 310 in figure \ref{fig8} which indicates  that the $\gamma$-ray absorption is negligible in the emission length scales here considered which are above $0.3 R_S$. 
 
 Also in this case  the   radiation in low energy range ($10-10^2 \rm{eV}$) is fully absorbed  due to photon-neutral interactions (Figure~\ref{icsed}).

As for M87, our model and the adopted parametrization can also  naturally explain the  fast variability of the VHE $\gamma$-rays in 3C 310.
 The effective emission zone for this source  is $L \simeq 0.3 R_S$ according to our model (see Table 1), which is compatible with the scale inferred  from the observed high variability in the $\gamma$-ray emission,   $\sim$ 4.8 min.

\section{Discussion and Conclusions}
 
Our main purpose here was to  explore an alternative mechanism to explain the spectral energy distribution (SED) and, particularly, the VHE emission in non-blazar AGNs, i.e., those for which the  jet probably does not point to the line of sight and therefore, do not have their emission enhanced by Doppler boosting in the relativistic jet.  Instead, we have examined an acceleration mechanism occurring in the innermost region of the AGN. Based on recent results by  \cite{beteluis2010a},  KGS15, and \cite{chandra14} we investigated the role of fast magnetic reconnection events in accelerating particles in the nuclear regions of low luminosity AGNs, applying this acceleration model to reconstruct the SED of Cen A, Per A, M87 and IC 310.  
 
According to this model,  trapped particles within the magnetic reconnection discontinuity formed by the encounter of  the magnetic field lines arising from the accretion disk with those of the BH magnetosphere (Figure 1), can be accelerated to relativistic velocities by a first-order Fermi process  in the surrounds of the BH, as described in \cite{gl05}. Magnetic reconnection events will occur specially when there is substantial  increase in the disk accretion rate which helps  to  press the two magnetic fluxes together.   Since turbulence is expected to be present in these systems (see \S. 2.1),  the reconnection  can be made naturally fast by it (\citealt{LV99} and \citealt{kowal09})

KGS15 and \cite{chandra14} used this model to compute the  magnetic power released by reconnection in the surrounds of the BHs and compared it with the core  radio and $\gamma-$ray luminosities of outbursts  of a sample with  more than 230 sources including the microquasars and LLAGNs of the fundamental plane of BH activity (\citealt{merloni})  and also hundreds of  blazars and GRBs. They have found that the reconnection power is  large enough and is correlated with the observed  luminosities of  the microquasars and LLAGNs, following the observed trend   for more than $10^{10}$ orders of magnitude in luminosity and mass of these sources. \footnote{We note that this power is  generally not enough to reproduce the radio or gamma-ray luminosities  of the blazars and GRBs, but this  is compatible with the notion that the  emission in these cases is produced  outside the core, at the jet that points to the line of sight and screens any deep nuclear emission.} 

These correlations found with the emission of microquasars and LLAGNs are very important, because  they connect the non-thermal emission, specially the VHE one,  with the core of these sources, offering a reliable, self-consistent mechanism for their origin. However, in order to determine the real effectiveness of this mechanism,   it is necessary to reproduce the SEDs of these sources based on this model. This has been done recently for two microquasars Cyg X1 and Cyg X3,  for which the observed SEDs have been successfully reproduced by the model above (KGV15). In this work we have  extended this  analysis applying the reconnection model to the four radio galaxies with detection of VHE emission up to TeV.

Considering all relevant leptonic and hadronic radiative loss times due to the interactions of the accelerated particles  with  matter, radiation and magnetic field    
in the core regions of Cen A, Per A, M87 and IC 310,  we compared these times with the acceleration times 
and found larger energy cut-offs for particles being accelerated by magnetic reconnection than by diffusive shock (see Figures~\ref{fig2}, \ref{fig4},  \ref{fig6}, and \ref{ic}). 
This result stresses  the importance of magnetic reconnection as a potential acceleration mechanism in the core regions around BHs  and compact sources in general.

 Moreover, we note that the maximum energies for the electrons and protons obtained from these comparisons, 
 are actually much smaller than the maximum possible energy that the particles can attain within the acceleration region in the reconnection sheet. The latter is constrained by the thickness of the acceleration region, i.e., $\Delta R_X$ (given by Eq.~\ref{3}) which must be larger than or equal to the particle Larmor radius, $r_L=E/ceB$. This implies that the maximum energy to which  the particles  can be accelerated by magnetic reconnection is  $\sim 10^{20}\ \rm{eV}$, so that the reconnection layer is large enough to accelerate the particles to UHEs. This  value  reassures the efficiency of this   acceleration  process  and suggests that if the surrounds of the BHs in AGNs  were not so full of interacting matter and radiation fields  they  might be excellent sites for the production of UHECRs.

The cut-off  values above were employed in the determination of the accelerated
particle fluxes in the construction of the SED of the sources.
The electron synchrotron cooling rate is found to be larger than any other loss mechanism in all leptonic energy range (Figures~\ref{2a}, \ref{4a}, \ref{6a}, and \ref{ica}). Also, it is the dominant process  providing  the  radiation field that  produces the SSC and the photo-meson ($p\gamma$) radiation in the SEDs of the  four sources. 
We have also found that  the $p\gamma$ process is the dominant mechanism to cool the accelerated relativistic protons in the high energy branch. This is shown in Figure~\ref{2b} for Cen A for proton energies $\geqslant 6\times10^{16}\rm{eV}$, and in Figure~\ref{4b} for Per A  for proton energies $\geqslant 2\times10^{15}\rm{eV}$, while   in figures~\ref{6b} and \ref{icb} for M87 and IC 310, we see that synchrotron radiation can cool the protons for energies $\leqslant 2\times10^{15}\rm{eV}$, but for larger energies it is overcome by the  $p\gamma$ mechanism.

In summary, we  have shown  that, employing  fiducial parameters, our acceleration model  is capable of explaining the  non-thermal low and high energy emission of the SED  of the four investigated LLAGNs Cen A (Figure \ref{3a}), Per A (Figure \ref{5a}), M87 (Fiugre \ref{7a}) and IC 310 (Figure \ref{icsed}). 

An interesting  advantage of the model presented here is the relatively small number of free parameters used to construct the SED (seven). 
In particular, we adopted an one-zone approximation in order  to avoid the introduction of more free-parameters. Furthermore, this work is a first attempt to test the acceleration model by magnetic reconnection in the surrounds of these BH sources whose physics and distribution of the photon, density and magnetic fields are still  poorly known making  the use of multi-zone models even more challenging or artificial. 
Unlike other models (e.g. \citealt{kachel09b}), our acceleration mechanism was used to constrain the characteristics of the acceleration region. While most of the models take  the maximum energy of the  particles as a free parameter to fit the SED (e.g. \citealt{abdo09c,abdo09a,abdo09,abdo10,kachel10}), our model determines this directly from the acceleration model, as described.

\subsection{Cen A}
According to our results for  Cen A, the observed hard X-rays and $Femi-LAT$  $\gamma$-ray data can be  interpreted as due to  SSC and $pp$ interactions, respectively, with    accelerated particles injected   in the nuclear region (at distances  $\sim 20 R_S$) driven by magnetic reconnection with a distribution with  power law index $p=2.4$.  The \rm{TeV} radiation observed by $HESS$, on the other hand, is explained by neutral pion decays resulting from $p\gamma$ interactions. 

In \citealt{sahu12}, the authors also showed that the \rm{TeV} $\gamma$-rays in Cen A could be explicated by $p\gamma$ interactions, but between relativistic protons  accelerated by  Fermi process in shocks along the jet with the monochromatic photons observed at $170\rm{keV}$. 
Another  model (\citealt{reynosoetal2011} also proposed particle acceleration  at the jet basis with the production of the hard X-rays by synchrotron emission, and the $Fermi-LAT$ and $HESS$ data  by IC and $pp$ mechanisms, respectively, along the jet.
\citealt{kachel10}, on the other hand, have argued against the production of the $\gamma-rays$ in Cen A  by   $pp$ interactions  along the jet because  on leaving the source they would interact with the EBL resulting in a flatter spectrum in the TeV range than the observed one by HESS 
(see also \S. 1). 

All these studies demonstrate that the origin of the VHE emission in this source is still highly debatable. A core origin, as the one suggested in this work arises as an interesting possibility as  long as magnetic activity is significant in the surrounds of BHs, and it might be considered as well. To disentangle this puzzle we will  need substantially improved observations, specially in the $\gamma-$ray range. It is possible that with the much larger resolution and sensitivity of the forthcoming CTA observatory  (Actis et al. 2011;Acharya et al. 2013; Sol et al. 2013),   and  with longer times of exposure of this nearby source (and also of  M87, Per A and IC 310), we may collect higher resolution data, and more significant information on variability that may help to determine the location of the emission region.

\subsection{Per A}

In the case of Per A, there is no relevant data yet  in  \rm{TeV} energies, but  our core model can nearly  explain the observed  $Fermi-LAT$ and $MAGIC$ data in the 0.1 GeV $-$ 650 GeV range with a leptonic scenario dominated by SSC. The synchrotron photons that are absorbed in SSC are produced by accelerated  electrons by magnetic reconnection  in the  coronal nuclear region around the BH (within  distances  $\sim 20 R_S$) having  a distribution with a power law index $p=2.15$.

An SSC model has been also proposed by \cite{aleksic14}, but they assumed  
 that the Per A core could be a BL Lac blazar with the jet bending strongly at larger scales  and the high energy non-thermal  radiation could be originated in a sub-structure of the jet near the core  pointing  towards our line of sight. 
This bending still requires observational support and any evidence of jet precession (e.g., \citealt{walker94,falceta10}) may favour this model. But  our proposed model dismisses the necessity of  such a strong bending and besides, is supported by the correlations with the observations  found  in KGS15 and \cite{chandra14}.


\subsection{M87}
 
In the case of M87, we have applied the same magnetic reconnection model in the nuclear region around the BH (within a region of 20 $R_S$) considering the injection of the accelerated particles with a power law index $p=2.4$. This has resulted  a lepton-hadronic scenario for the SED with SSC emission and neutral pion decays from $pp$ collisions explaining the $Fermi-LAT$ data. 
We also found that the decay of the neutral pions due $p\gamma$ interactions can explain the observed data by   $HESS$ in the TeV range. 

The suggested sites of TeV  emission  for this source in former works range from large scale structures of the kpc jet (\citealt{stawarz05}) to a compact peculiar hot spot (the so-called HST-1 knot) at a distance 100 pc along the jet (\citealt{stawarz06}) and inner (sub-parsec) parts of the sources. \cite{reynosoetal2011} for instance, considered that this  emission  is  produced in the jet, but   reconstructed all the emission features, which are  highly variable and possibly non-simultaneous,  with a single $pp$ mechanism.  
 
 \cite{gian10}, on the other hand,  proposed that compact minijets induced by magnetic reconnection moving relativistically within the jet in different directions, some of which pointing to  our line of sight,  might explain the short-time  variable TeV flares observed in M87. This model bears several similarities with ours as  it proposes that the minijets are generated by reconnection events in the core region,  and then move out with the jet flow up to scales of $\sim 100 R_S$. Our model also predicts the development of outbursts with the formation of reconnected features (plasmons) that may be carried out with the jet and might explain e.g.,  observed superluminal features  near the jet basis (\citealt{gl05}, \citealt{beteluis2010a}). However, \cite{gian10}  study  provides no predictions for the SED structure of M87.  

In addition, there is   an extensive list of  models that propose  that the variable VHE emission of M87  can be originated in the  inner jet. These span from leptonic  models, such as the decelerating flow (\citealt{georga05}), the spine-shear (\citealt{tavecchio08}), and the mini/multi-blob model (\citealt{lenain08}),  to hadronic models with the emission due to proton synchrotron-$p\gamma$ interactions (\citealt{reimer04}),  or  $pp$ interactions in a  cloud-jet scenario  (\citealt{barkov12}).  However, the location of the emission region is still an unsolved problem. 

 \cite{neronov07} also proposed a nuclear origin for the  TeV emission of M87 coming directly from the magnetosphere of the black hole (see also \citealt{levinson}). They showed that accelerated electrons in the strong rotation-induced electric fields in vacuum gaps in the BH magnetosphere, similar to a  pulsar magnetosphere,  could lead to  the observed TeV emission.  Since the acceleration and emission mechanisms occurs in a very compact region close to the event horizon of the BH, it potentially can explain the observed variability of TeV $\gamma$-ray emission from M87. Besides, as in our model, they also explain this emission as due to  $p\gamma$ interactions with an IR compact target photon field produced by synchrotron emission.  However, as stressed in \S. 3, the attenuation of $\gamma$-ray emission due to electron-positron pair production may be significant in distances smaller than or equal to $\sim R_S$ (Figure~\ref{fig8}), which may affect their results. In our model, the emission scales are larger ($\sim 5 R_S$) making these attenuation effects negligible.

\subsection{IC310}
In the case of IC 310, also a lepton-hadronic model in the nuclear region  is able to explain the observed SED features with protons and electrons accelerated by magnetic reconnection and injected in the emission region with a power law distribution with index p = 1.7. As remarked, the observed radio emission is well fitted by synchrotron  and the VHE  emission detected by \textit{MAGIC} can be explained by decays of neutral pions resulting from $pp$ and $p\gamma$ interactions. 

Our model with an appropriate choice of parameters is also able to explain naturally the  time variability detected in the sources here investigated. In particular, the very fast variability  reported  for the IC 310 $\gamma-$ray emission  of about $\sim 4.8$ min (\citealt{aleksic14b}) implies an emission region scale  of $\simeq 0.3 R_S$. 
To explain this variability and compactness of the emission region, \citealt{aleksic14b} suggested  that the  particles could be accelerated by  electric fields  in the BH  magnetosphere, as in pulsar models. Nevertheless,  as  demonstrated, the model described here reproduces the observed SED with an emission region with a similar size as above.

In conclusion, in the  construction of the  SEDs for the sources discussed here (Cen A, Per A, M87 and IC 310) based on our magnetic reconnection model in the core region, the observed emission at  low energies (radio to optical)  
 can be explained by synchrotron emission.
 SSC  with  target photons coming from electron synchrotron emission is the dominant (leptonic) mechanism to produce the observed hard X-rays and low energy $\gamma-$rays, while  neutral pion decays resulting from $pp$ inelastic collisions is the dominant hadronic process to produce the high energy $\gamma-$rays, and  neutral pion decays resulting from  photo-meson interaction ($p\gamma$) the dominant one to produce the very high energy (VHE) $\gamma-$rays.
 Interestingly, in the case of the microquasars Cyg X1 and Cyg X3, KGV15 have also found that the core model could reproduce the full observed SED including the low and high energy branches.
 \footnote{We should  remark that the observed emission at  low energies (radio to optical) from the core regions in the case of Cen A and M87 is   fitted by the core model described here only if we assume that the minimum electron energy injected  in the acceleration region is a few times the particle rest mass.
If one considers instead, $\gamma_{min} =2$ in the computation of the SEDs of these sources, the calculated synchrotron spectrum  produced inside the core, mostly in the IR band,  is fully absorbed by the energetic electrons and protons in order to  produce the SSC and $p\gamma$ emissions, respectively, at the higher energies. Thus if this were the case,  the observed radio to optical spectrum in these sources would be probably due to  more evolved synchrotron radiation  produced beyond the VHE emission region, probably in the jet basis, which would be  compatible with  jet-like models  for the low energy range (e.g., \citealt{reynosoetal2011}).}

\subsection{Conclusions}

\begin{itemize}
  \item We have presented a reconnection acceleration model in the core region around the BH of the low luminous radio galaxies Cen A, Per A, M87 and IC 310 and showed that it is able to reproduce very well their SEDs,  from radio to gamma-rays up to TeV energies. 

\item This is  complementary  to a recent study where we have performed similar analysis for the galactic black hole binaries, i.e., the microquasars Cyg X1 and X3 (Khiali et al. 2015). Together, these two works strengthen the conclusions of the previous works of Kadowaki et al. (2015) and Singh et al. (2015) in favour of a core emission specially for the observed  gamma-ray radiation of microquasars and LLAGNs (which belong to the so called fundamental plane of BH activity).

\item Magnetic reconnection acceleration seems to provide  a better efficiency  in regions where magnetic activity is dominant in comparison with  diffusive shock acceleration as  the cores of LLAGNs. Particles can gain energy up to a few times $\sim$100 TeV due to magnetic reconnection acceleration.
 
\item The observed TeV $\gamma$-ray emission may be originated in these cores via neutral pion decays in hadronic processes.
   

\item The fast magnetic reconnection acceleration model occurring in  the core of these sources can naturally  explain the observed short time variability, specially of the high energy $\gamma$-ray.

\end{itemize}

Finally,  we note that it is  possible that a neutrino spectrum may be also produced in the nuclear region of LLAGNs considering the same model here investigated, as due to charged pion decays via   $pp$ and $p\gamma$ interactions. In a concomitant work, this possibility has been investigated to explain the recently observed  extragalactic neutrino flux  by the IceCube experiment (\citealt{khiali15}).

\section*{Acknowledgements}
This work has been partially supported by grants from the Brazilian agencies FAPESP (2013/10559-5), CNPq (306598/2009-4), and CAPES.  We also acknowledge useful discussions with Paola Grandi in the selection of data for the construction of the SEDs.


\label{lastpage}

\end{document}